\documentclass[prx,twocolumn,reprint,amsmath,amssymb,letterpaper,longbibliography,footinbib]{revtex4-1}
\usepackage{amsfonts} 
\usepackage{graphicx}
\usepackage{dcolumn}
\usepackage{float}
\usepackage{bm}
\usepackage{mathrsfs}
\usepackage{txfonts}
\usepackage{tabularx}
\usepackage{bigstrut}
\usepackage[usenames,dvipsnames]{color}
\usepackage{natbib}

\newcommand{\com}[1]{}

\newcommand{\bS}{\mathbf{S}}

\newcommand{\rr}{\mathbf{r}}

\newcommand{\bsigma}{{\boldsymbol\sigma}}

\newcommand{\z}{{\hat z}}
\newcommand{\ve}{{\varepsilon}}

\newcommand{\beq}{\begin{equation}}
\newcommand{\eeq}{\end{equation}}
\newcommand{\beqa}{\begin{eqnarray}}
\newcommand{\eeqa}{\end{eqnarray}}

\begin{document}

\title{
%{Electronic fractionalization in integer quantum Hall systems with chiral magnetic ordering\\
%{\em how about:} Anyons in spontaneous integer quantum Hall systems}
Anyons in integer quantum Hall magnets}

\author{ Armin Rahmani$^1$, Rodrigo A. Muniz$^{1,2}$,  and Ivar Martin$^{1,3}$ } 
\affiliation{$^1$Theoretical Division, Los Alamos National Laboratory, Los Alamos, NM 87545, USA \\ 
$^2$International Institute of Physics - UFRN, Natal, RN  59078-400, Brazil\\ 
$^3$Materials Science Division, Argonne National Laboratory, Argonne, IL 60439, USA}

\begin{abstract}
%We show that spontaneous integer Hall systems, which emerge due to the interplay of itinerant electrons and noncoplanar magnetic ordering, support intrinsic fractional excitations, i.e., the magnetic $Z_2$ vortices in these systems, whose topological stability is guaranteed by the structure of the order parameter, generically carry fractional electric charge.

%There are two main routes to electronic fractionalization: The first is realized in fractional quantum Hall systems and is due to electron-electron correlations, and the other when electrons interact with topologically-nontrivial background field configurations. 

Strongly correlated fractional quantum Hall liquids support fractional excitations, which can be understood in terms of adiabatic flux insertion arguments. A second route to fractionalization is through the coupling of weakly interacting electrons to topologically nontrivial backgrounds such as in polyacetylene. Here we demonstrate that electronic fractionalization combining features of both these mechanisms occurs in noncoplanar itinerant magnetic systems, where $integer$ quantum Hall physics arises from the coupling of electrons to the magnetic background. The topologically stable magnetic vortices in such systems carry fractional (in general irrational) electronic quantum numbers and  exhibit Abelian anyonic statistics. We analyze the properties of these topological defects  by mapping the distortions of the magnetic texture onto effective non-Abelian vector potentials. We support our  analytical results with extensive numerical calculations.

\end{abstract}

\maketitle

\section{Introduction}\label{sec:introduction}
The discoveries of the integer and fractional Quantum Hall (QH) effects have revolutionized condensed matter physics: the important concept of a topological invariant was introduced to explain the quantized Hall conductivity of the former~\cite{Thouless1982}, while the novel notion of topological order, i.e., a type of nonlocal order with no Landau symmetry-breaking and no local order parameter, was introduced to describe the latter~\cite{Wen1990}. Topological order goes hand in hand with exotic phenomena such as fractional charge and statistics~\cite{Wen1990}. While the strongly correlated, topologically ordered fractional QH systems indeed have fractional quasiparticles \cite{Laughlin1983, Moore1991}, their more traditional weakly correlated counterparts have quasiparticles with integer charge.

Here we show that anomalous integer QH systems, which can emerge even in the absence of external magnetic fields in frustrated magnets, provide a new playground for electronic fractionalization.  While in conventional integer QH systems (two-dimensional electron gas in a magnetic field), fractional charge can only be induced artificially, e.g., near superconducting vortices~\cite{Weeks2007}, we demonstrate that $intrinsic$ topological defects in the noncoplanar magnetic systems, which exhibit anomalous integer QH effect, naturally harbor excitations with fractional electronic quantum numbers. We also show that these defects exhibit an Abelian anynonic exchange statistics.
\begin{figure}[ht]
\includegraphics[width=\columnwidth]{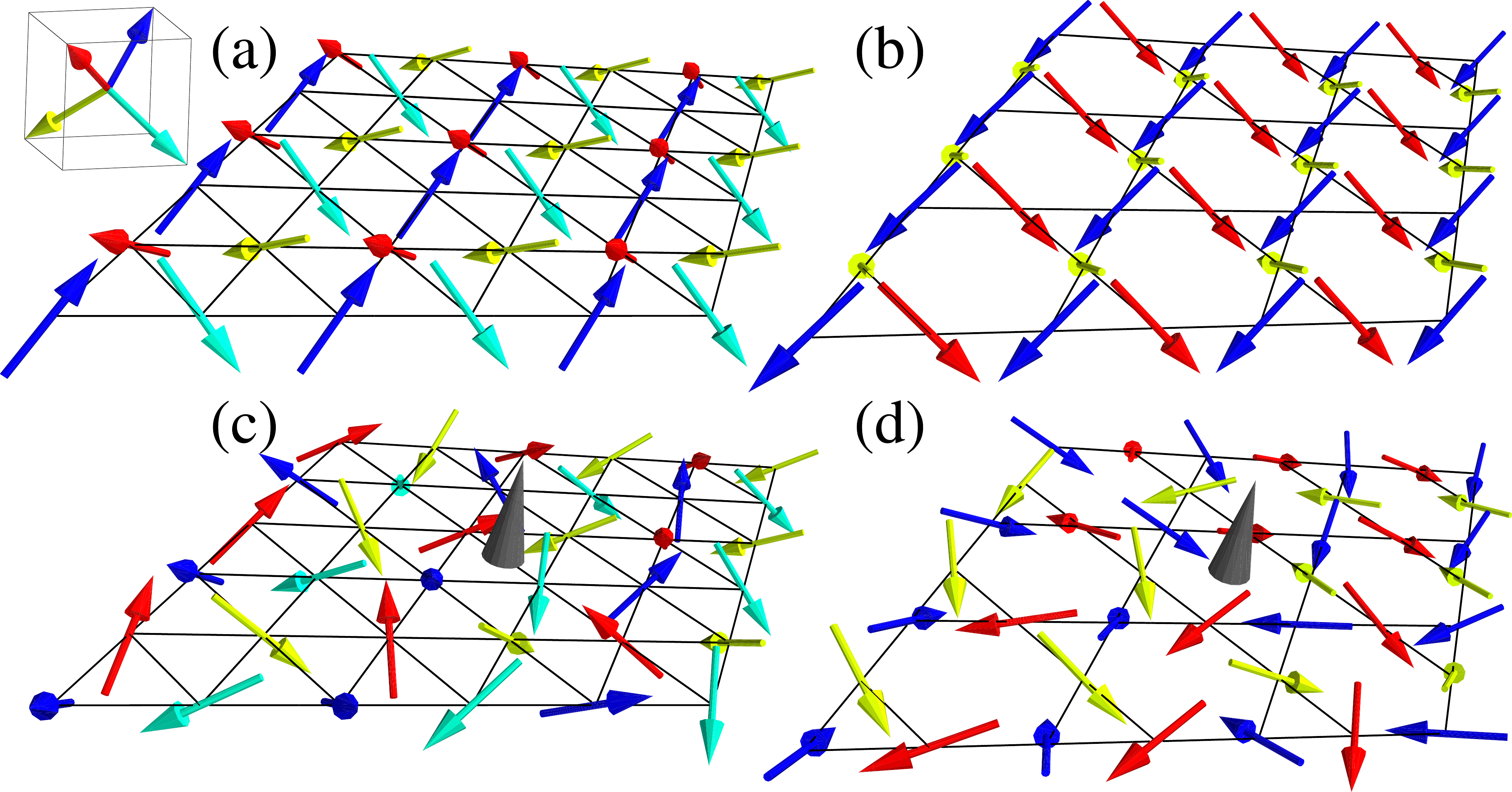}
\caption[]{(a) Chiral spin ordering on the triangular lattice. Four orientations of the local
magnetic moments correspond to the normals to the sides of a regular tetrahedron. (b) Chiral spin ordering on the kagom\'e lattice.
(c) Vortex configuration on the triangular lattice.  (d) Vortex configuration on the kagom\'e lattice. }
\label{FIG:1}
\end{figure}

The existence of fractional excitations in a QH liquid can be deduced from Laughlin's argument: upon adiabatic local ``insertion" of a flux quantum, a fractional charge $q = \sigma_{xy} e$, where $\sigma_{xy}$ is the fractional Hall conductivity (in units of  $e^2/h$), flows in from infinity~\cite{Laughlin1981,Laughlin1983}. Since integer flux quantum can be ``gauged away'' if its core is smaller than the physical unit cell size, the charge $q$ is in fact the charge of an elementary quasiparticle. Naturally, insertion of a fractional flux can also lead to the same result with an integer Hall response. Fractional fluxes, however, cannot be gauged away, and thus cannot be naturally associated with quasiparticles. It may thus appear that the existence of intrinsic fractional excitations in a QH system requires a fractional $\sigma_{xy}$.

An alternative route to electronic fractionalization is via interaction between electrons and topological defects in some order parameter, such as a pattern of lattice distortions in polyacetylene or graphene-like structures \cite{Jackiw1976,Su1979,Hou2007, Chamon2008,Seradjeh2008,Ryu2009}, a superconducting vortex \cite{Jackiw1981, Volovik1989, Read2000, Ivanov2001}, or a meron in a QH bilayer systems~\cite{Yang1994,Moon1995}. 
Here we analyze the scenario where QH effect itself emerges as a result of the coupling of itinerant electrons to chiral magnetic states. Such states can form through spontaneous time-reversal symmetry breaking \cite{Anderson1955, Haldane1988, Ohgushi2000, Taguchi2001,Shindou2001, Martin2008, Raghu2008},  as well as with an assistance of an external magnetic field that explicitly breaks the time-reversal symmetry.
Remarkably, the topologically stable defects in such ordered magnetic media act as effective fractional fluxes, giving rise to natural excitations with fractional charge. (In a broader context, studying topological defects in topological phases has attracted considerable recent interest \cite{Teo2010,Teo2013,Barkeshli2012,Barkeshli2013a,Barkeshli2013b}.)

The key ingredients of our model system are i) localized magnetic moments capable of forming a noncoplanar state that can be smoothly distorted at low energy cost, and ii) itinerant electrons which interact with the local moments, and possibly induce this state. Unlike collinear (such as the N\'eel) and coplanar (such as the ``120-degree") order, noncoplanar magnetic states are rarely stabilized for classical magnetic moments with short-range interactions (see Ref. \cite{Messio2011} for an exception). However, when magnetic moments are coupled to itinerant electrons, noncoplanar states are quite common, i.e., the magnetic phase diagrams of such itinerant systems seem to generically contain energetically stable phases with noncoplanar magnetic ordering~\cite{Ohgushi2000,Shindou2001,Akagi2010,Kumar2010,Martin2008,Kato2010,Li2011,Yu2012,Venderbos2012}. 
The interplay of noncoplanar magnetic moments and itinerant electrons can then lead to spontaneous quantized integer quantum Hall effect as a result of the nontrivial Berry phases imparted to the electrons by the noncoplanar magnetic texture \cite{Ohgushi2000, Shindou2001, Martin2008}. Such spin-chirality-driven Hall effect may be realized in a wide range of materials such as mangenites, CrO$_2$, the element Gd, the cobaltates, and pyrochlore ferromagnets (see Ref.~\cite{Nagaosa2010} and the references therein). To illustrate the generic nature of our results, here we consider two different models -- the triangular and kagom\'e Kondo lattices -- both exhibiting the same kind of  electronic fractionalization.

The outline of this paper is as follows. In Sec. \ref{sec:outline}, we summarize our main results. In Sec.~\ref{sec:model}, we briefly introduce the two models on the triangular~\cite{Martin2008} and the kagom\'e~\cite{Ohgushi2000} lattice. 
 In Sec.~\ref{sec:topology}, we discuss the topologically stable defects of noncoplanar magnetic states and give a general argument for charge fractionalization. In Sec.~\ref{sec:spinless} we present a microscopic theory of the fractional charge in limit of large Kondo coupling in the triangular lattice model, where the system maps to a simple model of spinless fermions. 
In Sec.~\ref{sec:numeric} we present numerical results for the properties of individual vortices, confirming the analytical calculations.
In Sec.~\ref{sec:berry}, we discuss the mutual exchange statistics of the vortices and present extensive numerical calculations, verifying our theoretical prediction of a relationship between the exchange statistics and the charge and magnetization of the vortices.

\section{ Summary of results }
\label{sec:outline}

Our results concern systems where local moments ${\bf S}_i$ form noncoplanar magnetic states, such as in Figs.~\ref{FIG:1}(a) and \ref{FIG:1}(b). When itinerant electrons interact with these magnetic states, for certain densities and ranges of interaction strength, they exhibit anomalous QH effect. The generic Hamiltonian that leads to this behavior has the form:
\beq
H =  -t_{ij} c_{i\alpha}^\dag c_{j\alpha}-\mu c_{i\alpha}^\dag c_{i\alpha}+ J   c_{i\alpha}^\dag \bS_i\cdot\bsigma_{\alpha\beta} c_{i\beta} + H_S\label{eq:H},
\eeq
where electrons hop on a two-dimensional (2D) lattice and interact with the local magnetic moments via onsite exchange interaction. Summation over repeated site (roman) and spin (greek) indices is implied. Here $t_{ij}$ is the intersite hopping (in this work we only consider nearest-neighbor hopping $t_{ij}$), $J$ is the exchange interaction constant, $\bS_i$ is the local magnetic moment (assumed classical), $\bsigma = (\sigma^x, \sigma^y, \sigma^z)$ is the vector of Pauli matrices, and $c_{i\alpha}$ is the operator of electron annihilation on site $i$ with spin $\alpha$. $H_S$ is the classical Hamiltonian that only includes spin variables.

The existence of topologically stable vortex defects in a noncoplanar magnetically ordered medium follows from the nontrivial fundamental homotopy group of the space of energetically degenerate configurations (order-parameter space). i) on the triangular lattice with $H_S=0$, the time-reversal symmetry can be broken spontaneously, leading to a spontaneous QH ground state (Fig. 1a.) for several electron densities \cite{Martin2008,Akagi2010,Kumar2010, Kato2010}. As the rotation of this texture around any axis keeps the energy unchanged, the the order-parameter space is the $SO(3)$ group. In this case,  the nontrivial fundamental homotopy group $\pi_1(O(3)) = \pi_1(SO(3)) = Z_2$ ~\cite{Kawamura1984} guarantees the existence of $Z_2$ vortices. ii) on the kagom\'e lattice with $H_S$ representing Heisenberg interactions between the local moments and their coupling to an external Zeeman field, the time-reversal symmetry is explicitly broken by the Zeeman field. In this model, the degenerate ground-state manifold has $SO(2)$ symmetry (as the texture can be rotated around the axis determined by the Zeeman field), and  $\pi_1(O(2)) = \pi_1(SO(2)) = Z$ results in vortices characterized by an integer winding number.

% {\color{red} 
%When the time-reversal symmetry is broken spontaneously, the order parameter space is the $SO(3)$ and the nontrivial fundamental homotopy group $\pi_1(O(3)) = \pi_1(SO(3)) = Z_2$ ~\cite{Kawamura1984} guarantees the existence of $Z_2$ vortices: this is exemplified by our model on the triangular lattice. 
%On the other hand, when the time-reversal symmetry is broken explicitly (by an external magnetic field or spin orbit coupling for instance), the order-parameter space is smaller. In the case for our model on the kagome lattice, we have an $SO(2)$ order-parameter space, and $\pi_1(O(2)) = \pi_1(SO(2)) = Z$ gives vortices characterized by an integer winding number.} 

An example of the vortex spin texture is shown in  Fig.~\ref{FIG:1}c and~\ref{FIG:1}d.
For instance, it can be obtained by rotating the order parameter (every magnetic moment)  by a position-dependent angle 
\beq \label {eq:phi}
\phi(\rr) = \nu   \arg (x + iy)
\eeq
around the $\z$ axis (assuming the vortex core is at the origin); $\nu$ is the winding of the vortex. In the triangular (kagom\'e) lattice case, the two topological classes set by the parity (value) of $\nu$. Unless stated otherwise, here we consider the $\nu = 1$ case.

Our main result is that the topologically nontrivial vortices have a fractional electric charge. If the vortex has a fixed axis of rotation $\hat n$ (in the kagom\'e lattice case above the axis is pinned to the magnetization direction), the value of the fractional charge is given by
\begin{equation}\label{eq:main0}
q=m\pm{1 \over 2}{ h \over e^2}\left[\sigma^{00}_{xy}+(\sigma^{0n}_{xy}-\sigma^{0n}_{yx})\right],
\end{equation}
where $m$ is an integer, and $\sigma^{00}_{xy}$ is the usual QH conductance, which characterizes the charge current flowing in the $x$ direction in response to an electric field in $y$ direction. Similarly, $\sigma^{0n}_{xy}$ characterizes the charge current flowing in the $x$ direction in response to a ``spin-$\hat{n}$ electric field" (to be defined later on) in the $y$ direction. For the textures on the triangular lattice without a uniform magnetization, the latter off-diagonal responses vanish, and for an integer QH system with $\sigma^{00}_{xy}=e^2/h$, the charge is given by
\begin{equation}\label{eq:main}
q=m+{1 \over 2},
\end{equation}
independently of the axis of rotation. 
 
 We further show that vortices with a fixed axis of rotation, carry a net magnetization $m_n$ in the direction of this axis, which stems from the spins of itinerant electrons. 
 We also demonstrate that vortices have anyonic exchange statistics, with a statistical angle that is related to both the charge and the magnetization of the defect, which for $q$ given by Eq.~\eqref{eq:main} is
\begin{equation}
\Theta=p\pi/2+\pi/4+\pi m_{n},
\end{equation}
where $p$ is an integer and $m_n$ is the aforementioned magnetization of the vortex. In addition to analytical arguments, we perform extensive numerical computations to test our results.

\section{Models and integer quantum Hall effect}
\label{sec:model}
The energy of classical configuration of magnetic moments in Eq.~\eqref{eq:H}
depends on the quantum electrons. The configuration of the magnetic moments can be thought of as a set of external parameters for a quadratic electronic Hamiltonian, which uniquely determines the electronic energy at zero temperature for a given filling fraction. 

On the triangular lattice with nearest neighbor hopping, it has been shown that the magnetic moments spontaneously form an ``all-out" noncoplanar texture around $1/4$ and $3/4$ filling fraction. At $3/4$ filling, this instability can be understood in terms of the perfect nesting of the Fermi surface with the three ordering wave vectors corresponding to the all-out state~\cite{Martin2008}. For general fillings the lowest energy states can be obtained by Monte-Carlo simulations~\cite{Akagi2010, Kumar2010, Kato2010}.
At the filling fractions that correspond to electronically gapped all-out state, the system exhibits an integer quantum Hall response. 

Similarly, integer QH appears on the kagom\'e lattice with the moments forming an umbrella-like state where the spins are canted away from a 120-degree ordered state~\cite{Ohgushi2000}, at filling fractions $1/6$, $2/6$, $4/6$ and $5/6$. 
This state can be either induced by an external magnetic field in magnetic systems with the nearest-neighbor antiferromagnetic exchange interaction, or can emerge spontaneously, due to spin-orbit coupling.
Both ways correspond to specific forms of the classical term $H_S$ in Hamiltonian~\eqref{eq:H}. In the case of nonzero external magnetic field, which we consider here, $H_s$ has a $SO(2)$ degenerate ground-state manifold characterized by $\sum_{i \in \bigtriangleup}{\bf S}_i \propto {\bf H}$, where $\bf H$ is the magnetic field, and $\sum_{i \in \bigtriangleup}$  indicates a vector sum of the magnetic moments in a triangle. 
Classically, order by disorder selects coplanar states at infinitesimal temperatures~\cite{Zhitomirsky2002}. 
 By generating a large number of such coplanar states satisfying the above constraint, and computing the ground-state energy of the fermionic Hamiltonian, we verified that these generic configurations have a higher energy than the ``umbrella"-like noncoplanar state of Fig.~\ref{FIG:1} at zero temperature. This indicates that in the presence of itinerant electrons, this noncoplanar state is selected out of the classical degenerate manifold. In the following we therefore  assume that the spins form an umbrella state with the canting angle $\arctan\left({1\over 2\sqrt{2}}\right)$ with respect to the plane of the lattice (chosen so that, as shown in Figs.~\ref{FIG:1}a and \ref{FIG:1}b, the three magnetic moments point to three corners of a tetrahedron similarly to the all-out structure of the triangular lattice case).

% Similarly, integer QH appears also on the kagom\'e lattice where the moments form a canted 120-degree ordered structure~\cite{Ohgushi2000}, at filling fractions $1/6$, $2/6$, $4/6$ and $5/6$. 
% Such state is stable in a magnetic field perpendicular to the plane of the lattice. The classical spin Hamiltonian $H_s$ has a degenerate ground-state manifold characterized by $\sum_{i \in \bigtriangleup}{\bf S}_i \propto {\bf H}$, where $\bf H$ is the magnetic field, and $\sum_{i \in \bigtriangleup}$  indicates a vector sum of the magnetic moments in a triangle. Classically, order by disorder selects coplanar states at infinitesimal temperatures~\cite{Zhitomirsky2002}. 
% By generating a large number of such coplanar states satisfying the above constraint, and computing the ground-state energy of the fermionic Hamiltonian, we verified that all these configurations have a higher energy than the ``umbrella"-like noncoplanar state of Fig.~\ref{FIG:1} at zero temperature. We thus believe that in the presence of itinerant electrons, this noncoplanar state is selected out of the classical degenerate manifold. 
% {\color{red} Notice that, unlike the triangular lattice model above, the time reversal symmetry in this kagome model is broken explicitly rather than spontaneously.}

\begin{figure}[ht]
\includegraphics[width=0.7\columnwidth]{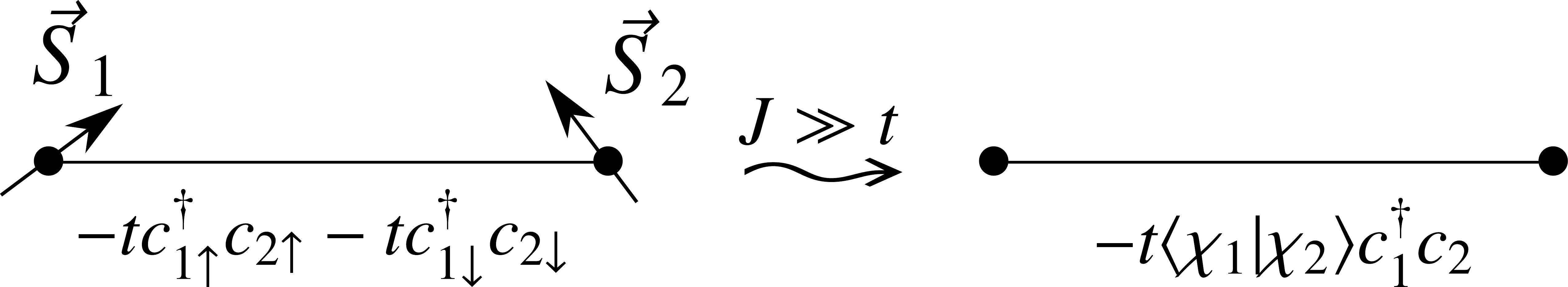} 
\caption[]{The projection of the Kondo-lattice model onto a spinless hopping model. Here $\chi_{i}$
is a spinor in the direction of $\vec{S}_{i}$, i.e., $\vec{S}_{i}\cdot{\boldsymbol{\sigma}}|\chi_{i}\rangle=|\chi_{i}\rangle$.}
\label{fig:proj} 
\end{figure}

The integer quantum Hall response in both cases described above stems from interaction between electrons and the noncoplanar magnetic state. Electrons hopping in a noncoplanar magnetic state are subject to an effective Berry phase, which results in a gapped integer quantum Hall liquid. 
This can be understood most simply in the limit of large exchange coupling $J$ 
\footnote{ Even though at infinite $J$, the noncoplanar structure may not be energetically stable, the Hamiltonian is adiabatically connected to the one with the same noncoplanar texture and finite $J$ (a regime where the texture is stable) and has the same topological properties.}.   
In this limit, the problem can be projected to a spinless hopping model, with each spinless electron representing an electron that is spin-polarized in the direction of the local magnetic moment.
The (gauge-dependent) hopping amplitude can be constructed as shown in Fig.~\ref{fig:proj}. Even though each individual hopping phase is gauge-dependent, the flux 
\[
\Phi_{123}=\arg\left[\langle\chi_{1}|\chi_{2}\rangle\langle\chi_{2}|\chi_{3}\rangle\langle\chi_{3}|\chi_{1}\rangle\right]
\]
through every triangular plaquette is gauge-invariant, and is given by half the solid angle subtended by the three magnetic moments $\vec{S}_{i}$~\cite{Fradkin1991}.
This flux is generically nontrivial for noncoplanar textures: in the particular case of Ref.~\onlinecite{Martin2008} (see Fig.~\ref{FIG:1}), e.g., we have a flux $\pi/2$ through each triangular plaquette.

\begin{figure}[ht]
\vspace{4mm}
 \includegraphics[width=\columnwidth]{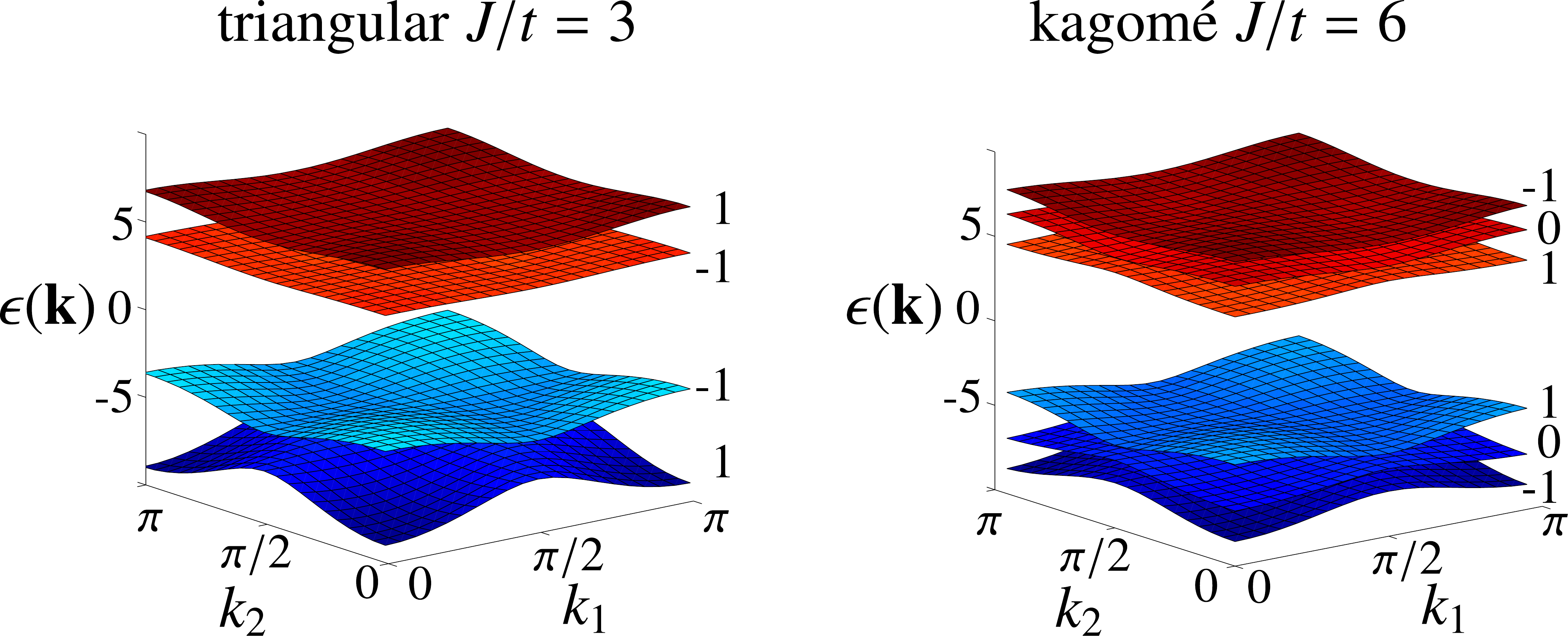}
\caption[]{The band structure of Hamiltonian [Eq.~\eqref{eq:H}] for the triangular and kagom\'e lattice model with textures of Fig.~\ref{FIG:1}. The Chern numbers of the individual bands are indicated to their right. Here $\mathbf{k} =k_1 \mathbf{Q}_1+k_2 \mathbf{Q}_2$, with $\mathbf{Q}_1=(1,-1/\sqrt{3})$ and $\mathbf{Q}_2=(0,2/\sqrt{3})$.}
\label{fig:band} 
\end{figure}

This Berry phase has a similar effect to an external magnetic field and gives rise to bands with nontrivial Chern numbers. As seen in Fig.~\ref{fig:band}, the spectrum of the triangular lattice model consists of four doubly degenerate bands with Chern number $+1$, $-1$, $-1$, and $+1$.
Similarly, as seen in Fig.~\ref{fig:band}, the spectrum of the kagom\'e lattice model \cite{Ohgushi2000} consists of six bands with Chern number $-1$, $0$, $+1$, $+1$, $0$ and $-1$.

\section{Topological defects and fractionalization}
\label{sec:topology}

\subsection{Topological defects from homotopy theory}

\begin{figure}[ht]
\includegraphics[width=0.9\columnwidth]{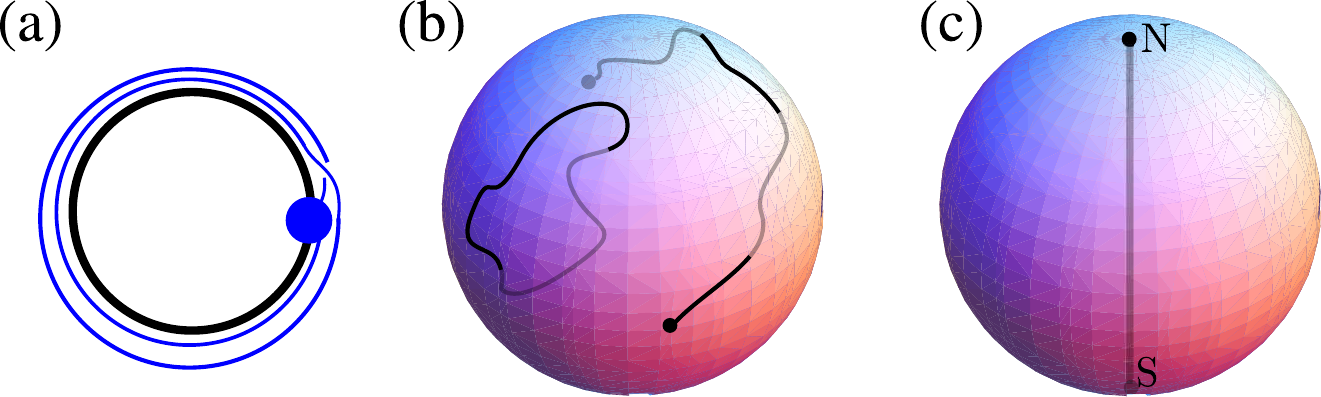}
\caption[]{(a) The $SO(2)$ order-parameter space and a vortex of winding number 2. (b) The $SO(3)$ order-parameter space and the two types of topologically distinct loops: a trivial loop and a vortex. The loops can lie on the surface of (black) or inside (gray) the solid sphere. The loops corresponding to a vortex connect two antipodal points on the surface of the sphere. (c) The nontrivial loop corresponding to the vortices shown in the Figs. 1(c) and 1(d).}

\label{FIG:so3sphere}
\end{figure}

As mentioned before, the spin-rotational symmetry can be broken either explicitly or spontaneously, which affects the structure of the degenerate manifold (``the order parameter space").
%
% In the systems above, time-reversal symmetry is broken both by the magnetic ordering  $\langle \bS_{i}\rangle $ and by the scalar chirality $\chi_{ijk}=\left\langle \bS_{i}\cdot \bS_{j}\times \bS_{k}\right\rangle $, where $i$, $j$ and $k$ represent the three sites in a triangular plaquete. The sign of chirality determines the sign of the Hall conductance $\sigma_{xy}$. Let's focus here on one chiral domain with a fixed $\sigma_{xy}$. 
%
In our kagom\'e lattice model, the out-of-plane direction of the magnetic field is fixed, but all the magnetic moments can be simultaneously rotated around this axis while preserving energy. Therefore the order parameter space is given by $SO(2)$, which can be geometrically represented by a unit circle. A loop in real space then maps to a loop in the order-parameter space, giving rise to regular $SO(2)$ vortices, which as illustrated in Fig.~\ref{FIG:so3sphere}(a), are characterized by an integer winding number $\pi_1(SO(2))=Z$.

On the other hand, in the triangular lattice model there is no such preferred axis in the absence of an external Zeeman field or spin-orbit coupling. 
Then the order-parameter space corresponds to a full rotation matrix in 3D, which can be parametrized by an angle and an axis. This space can be geometrically represented by a solid sphere of radius $\pi$ with antipodal points on the surface identified: the distance from each point to the center of the sphere represents the angle of rotation, while the vector connecting the point to the center gives the axis of rotation. 
The identification of antipodal points follows from the fact that clockwise and counterclockwise rotations around the same axis by angle $\pi$ are equivalent. A 1D loop in real space maps onto a 1D loop in the order-parameter space, which as seen in Fig.~\ref{FIG:so3sphere}(b) can fall into two distinct topological categories: contractible (topologically trivial), and noncontractible (topologically nontrivial). Mathematically, this classification is encoded in the fundamental homotopy group of the 3D rotations $\pi_1 \left( SO (3) \right) = Z_2$ \cite{Mermin1979}.

The noncontractible loop  corresponds to a nontrivial vortex. For example, the vortex configuration of Fig.~\ref{FIG:1}(d), which has a well-defined axis of rotation in the $\hat{z}$ direction, corresponds to the following noncontractible loop: a straight line passing through the North pole, the center of the sphere and the South pole (which is identified with the North pole) as shown in Fig.~\ref{FIG:so3sphere}(c).

\subsection{Fractionalization from Laughlin's argument}

Vortices are inhomogeneities in the magnetic texture, which  correspond to a position- and possibly time-dependent distortion of some reference state, $\bS_i = {\cal R}(\rr_i, t)\bS_i^0$. These inhomogeneous states can be mapped onto a state with homogeneous order parameter, but in the presence of an effective (in general) non-Abelian vector potential as follows. 
The rotation of the order parameter in the classical spin space can be transformed into a unitary rotation $U(\rr_i,t) \equiv U_i$ of the electron spinors, according to $U_i^\dag \bsigma\cdot\bS_i U_i = \bsigma\cdot\bS_i^0$. Introducing new fermions $\psi_i = U(\rr_i, t)^\dag c_i$ (spin index is suppressed), the Hamiltonian (\ref{eq:H}) becomes
\beqa
H = &- i \psi_{i}^\dag U_i^\dag \partial_t U_i \psi_{i}  -t_{ij} \psi_{i}^\dag  U_i^\dag U_j \psi_{j}\nonumber\\
&-\mu \psi_{i}^\dag  \psi_{i}+ J  \bS_i^0\cdot \psi_{i}^\dag \bsigma \psi_{i}+H_S\label{eq:Hrot}.
\eeqa
This mapping allows one to conveniently calculate the charge and spin currents in response to the order parameter distortions. In particular, the vortex configuration corresponds to a spatially localized non-Abelian flux, which can be used to determine the charge of the vortex.

Assuming that the variation of the texture is slow on the lattice constant scale, we can make an expansion,
$U_i^\dag U_j = U_i^\dag [U_i + (\rr_j - \rr_i)\cdot \nabla U_i]$. 
It is convenient to introduce $SU(2)$ vector potential ${\bf A^{\nu}} = - i U^\dag \partial^\nu U \equiv {\cal A}_a^\nu \sigma_a$, with the indices $\nu = \{t,x,y\}$ representing the space-time components and $ a = \{1,2,3\}$ the $SU(2)$ generators. The Hamiltonian can then be written as 
\beq
H = H_0 - J_a^\nu {\cal A}_a^\nu, \label{eq:HJ}
\eeq
where $H_0$ is the Hamiltonian corresponding to the static undistorted  spin state, i.e., Eq.~\eqref{eq:H} with the substitutions $c \rightarrow \psi$ and ${\bf S}_i\rightarrow {\bf S}_i^0$, and the currents are defined as
\beqa
J_a^0& =&\psi(\rr_i,t)^\dag \sigma_a\psi(\rr_i,t),\\ 
J_a^x &= & i t_{ij}(x_i - x_j) \psi(\rr_i,t)^\dag \sigma_a\psi(\rr_j,t), \label{eq:jx}\\
J_a^y &= & i t_{ij}(y_i - y_j) \psi(\rr_i,t)^\dag \sigma_a\psi(\rr_j,t).\label{eq:jy}
\eeqa
Denoting the $2\times 2$ unit matrix by $\sigma_0$, for $a=0$ the definitions above also give the charge density and current operators 
\footnote{The physical observables are defined in terms of the original $c$-fermions, and while charge density is the same in the old and the new bases, $c_i^\dag c_i = \psi(\rr_i)^\dag\psi(\rr_i)$, other operators, such as spin, need not be. Although the fields couple to currents similar to gauge fields, they represent physical distortions of the magnetic medium, and do not have $SU(2)$ gauge symmetry}.

What is the vector potential that corresponds to a vortex? The vortex texture is necessarily singular near the core; therefore, the transformation to unwind the vortex is singular as well. The simplest transformation that takes the vortex texture (with the $z$ axis as the axis of rotation) into a uniform one is  $e^{i\sigma_3 \phi(\rr)/2}$, where $\phi(\rr)$ is the angle of rotation [see Eq.~(\ref{eq:phi})] around the $z$ axis.
However, since upon going around the vortex, $\phi(\rr)\to \phi(\rr) + 2\pi$, the unitary changes sign, this would correspond to antiperiodic boundary conditions for fermions along a line connecting the vortex to infinity. 
To avoid this complication,  the above $SU(2)$ transformation can be augmented by a $U(1)$ one \cite{Abanov2000}. The combined transformation $U(\rr)= e^{i(\sigma_3+1) \phi(\rr)/2}$ is only acting on up-fermions. Its associated vector potential is
$${\bf A^\nu} =\frac{1+ \sigma_3}{2}\,\partial^\nu \phi,$$
which has the field strength zero everywhere except for the vortex core.

Due to the singular nature of the vector potential, we cannot directly apply the linear-response formalism in the vicinity of the vortex core. To calculate the vortex quantum numbers, we can instead invoke an analog of the Laughlin's argument~\cite{Laughlin1981,Laughlin1983}. The flux of the non-Abelian gauge-like field through the vortex core is 
\begin{equation}\label{eq:guage}
{\bf \Phi} = \oint {\bf A} d\rr = ({1+ \sigma_3})\pi.
\end{equation}
Now, suppose that the flux is turned on adiabatically from zero to $\bf \Phi$. That will generate a non-Abelian emf acting on electrons, which at large distances from the core will be nearly uniform (tangential to any circle centered  at the vortex). The vortex quantum numbers are then obtained by integrating the associated currents generated in response to this emf.

If the texture is slowly  varying in time (compared with the inverse energy gap in the spectrum), as well as in space, the vector potential ${\cal A}_a^\nu$ is small and the expectation values of the current operators defined in Eqs.~\eqref{eq:jx} and \eqref{eq:jy} can be calculated with the linear-response theory. The charge and spin current are related to the vector potential through 
$\langle J_a^\eta \rangle= {\sigma^{ab}_{xy}} \epsilon_{\eta\mu\nu}\partial^\mu {\cal A}_b^\nu$. The Laughlin argument yields the charge by integrating the current in a dynamical process where the vortex is created adiabatically (the lattice provides an underlying regularization). This argument relies on two conditions: First, we need to have a continuous sequence of \textit{gapped} Hamiltonians connecting the one with flux zero to the one with $\bf \Phi$. Second, we need a continuity equation relating the currents we can calculate in linear response to quantum number densities. We have explicitly identified a sequence of gapped Hamiltonians in the limit of large $J$ and we thus expect that an adiabatic process exists for an arbitrary $J$ as well. The quantum numbers of interest for us are charge and magnetization. Since total electron number commutes with the Hamiltonian, the charge current strictly satisfies a continuity equation and we can use the Laughlin argument to compute the charge quantum number. We do not have such continuity equation for the spin current, and thus our vortex defects do not have a well-defined spin quantum number that would be independent of the exact vortex configuration.

We are now in a position to state our main result [Eq.~\eqref{eq:main0}] for the fractional charge. If we have a fixed axis of rotation, say $\hat z$ as in Eq.~\eqref{eq:guage}, the charge current flowing toward the vortex core gets contributions from $\sigma_{xy}^{00}=-\sigma_{yx}^{00}$ and similarly from $\sigma_{xy}^{03}$ and $\sigma_{yx}^{03}$. Note that in general the underlying lattice could break rotation symmetry  and the last two response functions could be different. Using Laughlin's argument, we immediately obtain the second term in Eq.~\eqref{eq:main0}: An ``electric field'' $\cal E$ in the tangential direction gives a current $\vec{J}=(\sigma_{xy}{\cal E }\cos \theta,-\sigma_{yx}{\cal E} \sin \theta)$  at polar angle $\theta$. The current flowing toward the vortex core is then given by $J_r=-\sigma_{xy}{\cal E }\cos^2 \theta +\sigma_{yx}{\cal E} \sin^2 \theta$. Since the average of both $\sin^2 \theta$ and $\cos^2 \theta$ is $1/2$, Eq.~\eqref{eq:main0} (modulo the integer $m$) follows upon integration over $\theta$.

The origin of an undetermined additive integer can be understood as follows.  There are many possible choices of the single-valued gauge transformations that unwind the vortex; e.g., another choice could lead to the vector potential ${\bf A^\nu} =({-1+ \sigma_3})\,\partial^\nu \phi/2$.  While such different choice does not change the magnetization, the charge accumulated in the vortex core changes sign. This ambiguity is naturally understood in terms of the electron occupancy of a particular localized electronic state, $\ve_0$, inside the spectral gap. When this state is empty, the charge of the system is $q$, and when it is occupied, the charge is $q+1$, all relative to the uniform state. In general, there can be more than one localized state inside the vortex core. Occupying any of these states increases the vortex charge by one electron charge [this corresponds to more general choices, ${\bf A^\nu} =({1 + 2n + \sigma_3})\,\partial^\nu \phi/2$, with $n$ any integer]. In case when 
the order parameter space is $SO(3)$, as is the case for triangular lattice model in zero magnetic field, by direct calculation we can verify that  $\sigma_{yx}^{0a}$ vanishes for $a\neq 0$ and the charge of the vortex remains half-odd-integer for an odd vorticity.  On the other hand, for an even vorticity, the charge induced according to the Laughlin argument will be integer. For $Z_2$ vortices, this is consistent with the homotopy classification that says that double vortex can be  smoothly connected to a uniform state, and thus the charge of quasiparticle associated with the double vortex can only be integer.

Through a explicit calculation described in Appendix.~\ref{app:response}, we can compute the necessary linear-response functions. As stated before, for the triangular lattice model we find:
\beq\label{eq:sigma}
{\rm triangular:}\quad \sigma^{00}_{xy} =-\sigma^{00}_{yx} = -e^2/h, \quad\sigma^{0a}_{xy} = \sigma^{0a}_{yx}=0 ,\quad  a\ne 0
\eeq   
The signs of both conductivities are flipped by switching between $^3/_4$ and $^1/_4$ fillings, or by changing the sign of the chiral ordering. The kagom\'e lattice model, on the other hand, has net magnetization, and, hence, in addition to $\sigma_{xy}^{00}=-\sigma_{yx}^{00}=e^2/h$, the following off-diagonal responses are nonzero: 
\[
\text{ kagom\'e}: \quad \sigma_{xy}^{02}=\sigma_{yx}^{02},\quad\sigma_{xy}^{03}=-\sigma_{yx}^{03}.
\]
This results in the same $q=1/2$ charge for an axis of rotation in the $xy$ plane ($ \sigma_{xy}^{01}=\sigma_{xy}^{10}\approx 0$), but if the axis of rotation has a component in the $\hat z$ direction (the direction of the overall magnetization), we get nonuniversal $J$-dependent contributions to the fractional charge.

Additionally, the adiabatic creation of the vortex generates a spin current through nonvanishing responses such as $\sigma_{xy}^{aa}$ for $a\neq 0$. For example, in the triangular lattice model we find $\sigma_{xy}^{aa}={1\over 3} {e^2\over h}$. As we stated before, there is no continuity equation for spin.
 However, since the divergence of the induced current is zero far from the vortex core, it is expected that the spin current will be nearly conserved everywhere, except near the vortex core where spin density accumulates. This suggests that the \textit{magnetization} attached to a vortex might be still close to the expected value obtained from integrating the spin current. With the assumptions above, since the final flux both in spin-$\sigma_3$ and charge channels is half of the flux quantum, the accumulated magnetization for the triangular lattice may be close to
\footnote{Since the gauge transformation only involves $\sigma_3$ Pauli matrix, the spin density in this channel is identical both in $c$ and $\psi$ bases.}
$m_z \approx 1/12$. 
(the extra $1/2$ for $m_z\approx{1\over 2}\times {1\over 2}\times{1\over3}$ is due to the fact that electron spin is $\bsigma/2$.)
We will numerically examine this approximate result for $m_z$ in the subsequent sections, and find that it underestimates the average magnetization by up to $40\%$ due to spin nonconservation.

\section{Microscopic derivation for $J\to \infty$}

\label{sec:spinless}
In this section we present a direct microscopic derivation of the fractional charge in the limit of $J \to \infty$. We only consider the triangular lattice model for brevity. This calculation is illuminating as it provides a step-by-step derivation of the charge accumulation directly on the lattice. As stated before, due to the alignment of the electron spin with local moments, in this limit we have a model of spinless fermions coupled to $U(1)$ fluxes (for arbitrary values of $J$  we had spinful fermions coupled to an $SU(2)$ fluxes). Therefore, this calculation also sheds light on the mechanism for the emergence of fractional flux.

%Let us begin by the Kondo lattice model on the triangular lattice with the Hamiltonian
%\[
%H=-t\sum_{\langle i,j\rangle} c^\dagger_{i \alpha}c_{j \alpha}+{\rm H.c.}+J\sum_i {\mathbf S}_i\cdot c^\dagger_{i\alpha}{\boldsymbol\sigma}_{\alpha \beta} c_{i \beta}
%\]
%where $c_{i\alpha}$ is the fermion annihilation operator with spin $\alpha$ on site $i$, ${\mathbf S}_i$ is a classical magnetic moment on site $i$ and ${\boldsymbol\sigma}=(\sigma^x,\sigma^y,\sigma^z)$ is a vector of Pauli matrices. The coupling constants $t$ and $J$ are respectively the hopping amplitude and the exchange interaction constant and $\langle i,j\rangle$ indicates nearest neighbor. In the limit of large $J$, each itinerant electron aligns itself with the local magnetic moment and we can write an effective Hamiltonian for spinless fermions
The $t_{ij}\rightarrow t_{ij}=t\langle\chi_{i}|\chi_{j}\rangle$ mapping,
where $|\chi_{i}\rangle$ is a spinor in the ${\mathbf{S}}_{i}$
direction, modifies both the magnitude
and phase of the hopping amplitude. As we will see, however, the important physics stems
from the change in phase. The following effective
Hamiltonian then captures the physics of the problem in the large-$J$ limit: 
\begin{equation}
H=-t\sum_{\langle i,j\rangle}\frac{\langle\chi_{i}|\chi_{j}\rangle}{|\langle\chi_{i}|\chi_{j}\rangle|}c_{i}^{\dagger}c_{j}+{\rm H.c}.\label{eq:hamiltonian}
\end{equation}

\begin{figure}[ht]
\includegraphics[width=0.9\columnwidth]{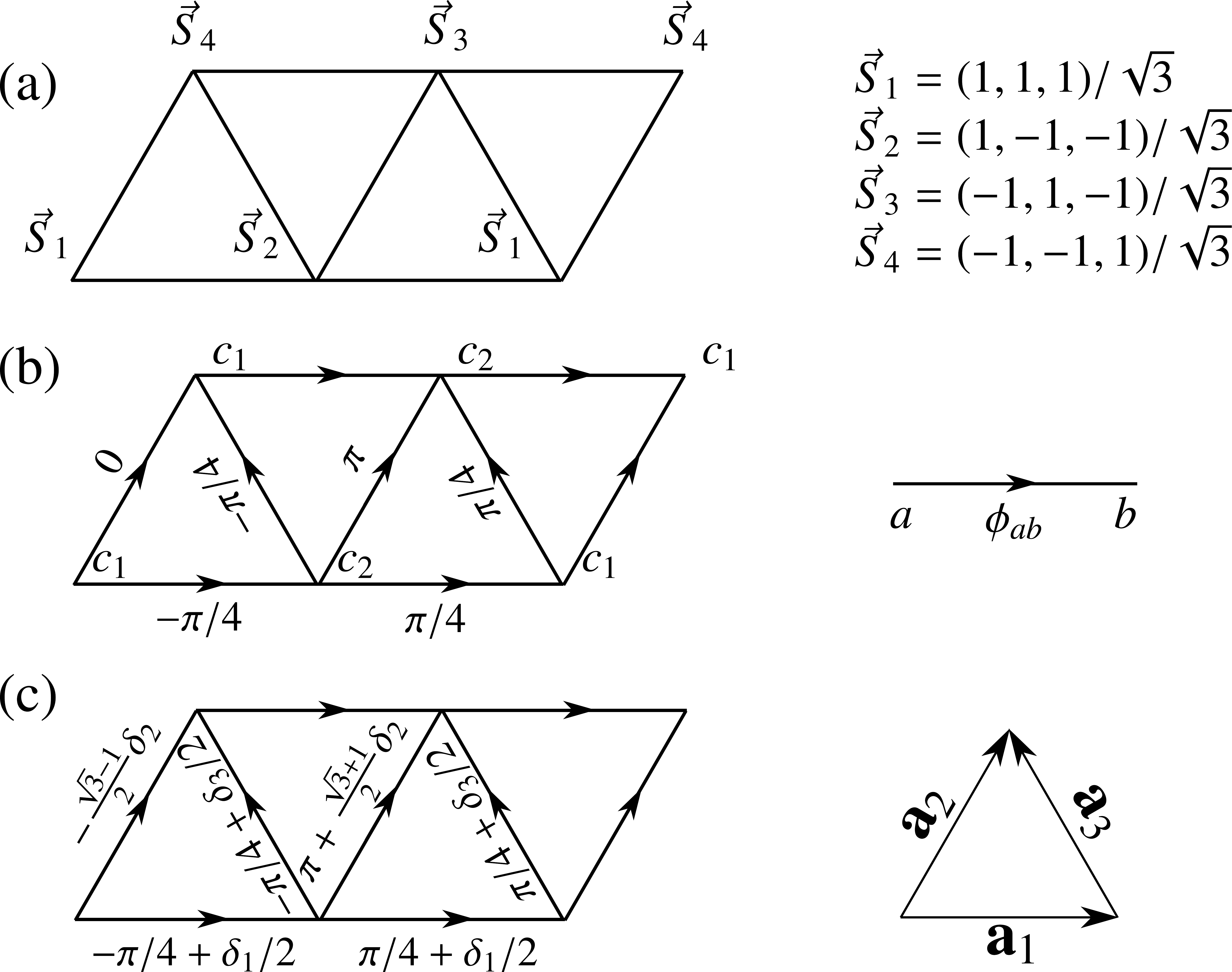} 
\caption[]{(a) spin components for the texture of Fig.~\ref{FIG:1}a. (b) The phases $\phi_{ab}$ of the hopping amplitude on different links
of the lattice. The hopping Hamiltonian can then be written as a $2\times2$ matrix in the basis represented by $c_{1,2}$. (c) The phases $\phi_{ab}$ of the hopping amplitude in the presence of a vortex [see Eq.~\eqref{eq:delta}], and the lattice vectors ${\bf a}_i$.}
\label{fig:trig} 
\end{figure}

We pick a coordinate system to represent the magnetic moments of Fig.~\ref{FIG:1}(a) (this choice is arbitrary as there is no spin-orbit coupling). Consider a chiral texture of magnetic moments on the triangular lattice as in Fig.~\ref{FIG:1}(a) [we have written out explicit spin components in Fig.~\ref{fig:trig}(a) below].
To find the hopping amplitudes
in Eq.~(\ref{eq:hamiltonian}), we need to calculate the spinors
$|\chi_{a}\rangle$ for $a=1\dots4$, which, in terms of the angle
$\beta=\arccos\left({1/\sqrt{3}}\right)$, are given by 
\begin{eqnarray*}
 &  & |\chi_{1}\rangle=\left(\begin{matrix}\cos\frac{\beta}{2}\\
\sin\frac{\beta}{2}\: e^{i\pi/4}
\end{matrix}\right),\qquad|\chi_{2}\rangle=\left(\begin{matrix}\sin\frac{\beta}{2}\\
\cos\frac{\beta}{2}\: e^{-i\pi/4}
\end{matrix}\right)\\
 &  & |\chi_{3}\rangle=\left(\begin{matrix}\sin\frac{\beta}{2}\\
\cos\frac{\beta}{2}\: e^{3i\pi/4}
\end{matrix}\right),\qquad|\chi_{4}\rangle=\left(\begin{matrix}\cos\frac{\beta}{2}\\
\sin\frac{\beta}{2}\: e^{-3i\pi/4}
\end{matrix}\right).
\end{eqnarray*}
We then obtain the phases $\phi_{ab}$, which are defined through
$\frac{\langle\chi_{a}|\chi_{b}\rangle}{|\langle\chi_{a}|\chi_{b}\rangle|}\equiv e^{i\phi_{ab}}$,
and are graphically represented in Fig.~\ref{fig:trig}(b).
%\begin{eqnarray*}
%& &\phi_{12}=-\pi/4,\qquad \phi_{23}=\pi,\qquad \phi_{43}=-\pi/4
%\\
%& &\phi_{31}=-\pi/4,\qquad \phi_{24}=-\pi/4,\qquad \phi_{14}=0.
%\end{eqnarray*}
%Setting the hopping amplitude $t$ to unity, we can then graphically represent the Hamiltonian Eq.~(\ref{eq:hamiltonian}) as in Fig.~\ref{fig:H0}.

In the basis $\Psi_{\mathbf{k}}^{\dagger}=\left(c_{1{\mathbf{k}}}^{\dagger}\quad c_{2{\mathbf{k}}}^{\dagger}\right)$,
with the annihilation operators $c_{1}$ and $c_{2}$ shown in
Fig.~\ref{fig:trig}(b), the Hamiltonian can be written as a $2\times2$
matrix $H_{0}({\mathbf{k}})$ in momentum space, i.e., $H=\sum_{\mathbf{k}}\Psi_{\mathbf{k}}^{\dagger}\: H_{0}({\mathbf{k}})\:\Psi_{\mathbf{k}}$.
\begin{figure}[ht]
\includegraphics[width=0.6\columnwidth]{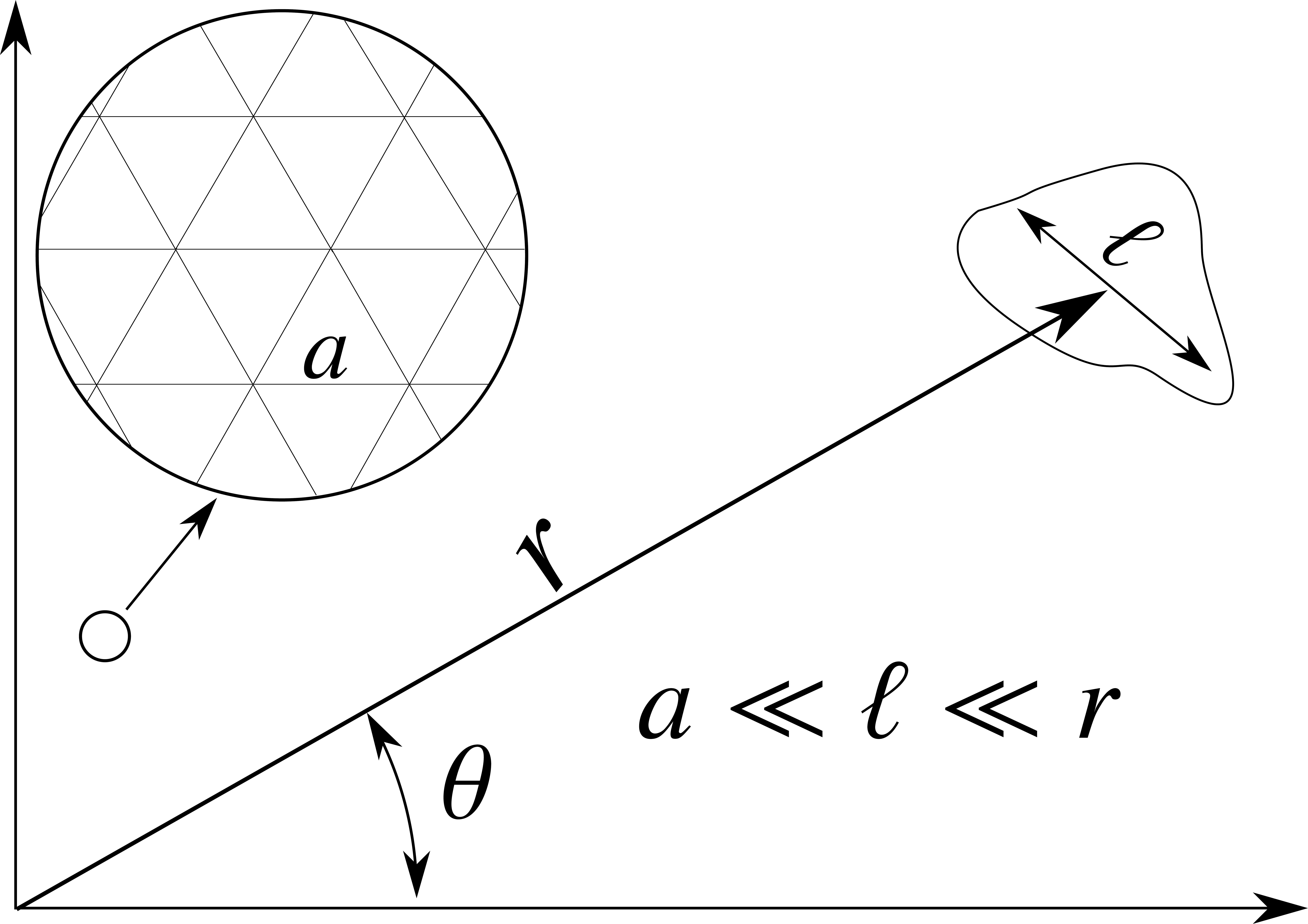} \caption[]{In a region of size $\ell$ far away from the vortex, the perturbation
Hamiltonian is translationally invariant.}
\label{fig:region} 
\end{figure}

We are interested in the effect of an $SO(3)$ vortex in the magnetic order on the itinerant electrons. Consider a vortex obtained by rotating each magnetic moment ${\mathbf{S}}_{i}$ around the $\hat{z}$ axis by an angle $\theta$ equal to the polar coordinate of site $i$ (in a planar polar coordinate system with the vortex core at the origin).
The rotation angles are not small but their difference for two nearby magnetic moments is small far away from the vortex:
it scales as $r^{-1}$. We have chosen a convenient gauge here such that the spin rotations due to the vortex give a small perturbation to
the Hamiltonian far away from the vortex core.

%This is because, in this gauge
%\[
%\arg\: \langle \chi_a|\chi_b\rangle=\arg \:\left[\langle \chi_a|\chi_b\rangle \langle \chi_b|\chi_\uparrow\rangle \langle\chi_\uparrow|\chi_a\rangle \right]
%\]
%where $|\chi_\uparrow\rangle$ is the spinor in the positive $\hat{z}$ direction, i.e. $\langle \chi_\uparrow|=(1,0)$. The right hand side of the above equation however is half the solid angle distended by $\hat{z}$ and the two moments $\vec{S}_a$ and $\vec{S}_b$. Now if these two magnetic moments are rotated by the same angle around the $\hat{z}$ axis, it it easy to see that the solid angle does not change. As we will see below, this convenient gauge allows the application of the linear response theory. 

Our approach is then to calculate the expectation value of the current
flowing toward the vortex core, in a region of size $\ell$ depicted
in Fig.~\ref{fig:region}, which is much larger than the lattice
spacing $a$ and much smaller than its distance $r$ from the vortex
core, while turning on the vortex adiabatically. Note that the lattice
provides short-distance regularizations so the singularity of the
vortex at the core is not an issue (unless the core is sitting right
on a lattice site). The condition $a\ll\ell\ll r$ allows us to treat
the problem in the region of size $\ell$ as approximately translationally
invariant (analogous to gradient expansion methods~\cite{Chamon2008,Santos2011}). Through a Laughlin-type argument, we can then find the
charge bound to the vortex by integrating this current.

The first step to carrying out the above procedure is to find the
first-order correction to $\phi_{ab}$ in the region shown in Fig.~\ref{fig:region}.
The correction for each bond depends on the difference between the
rotation angles of the magnetic moments at the two ends of the bond,
which we represent by $\delta_{ab}$. To leading order, in a region
labeled by $r$ and $\theta$ as in Fig.~\ref{fig:region}, $\delta_{ab}$
depends only on the direction of the bond. There are three types of
bonds corresponding to the three lattice vectors ${\mathbf{a}}_{i}$,
$i=1,2,3$ so we get three types of
$\delta_{ab}\equiv\delta_{i}$, $i=1,2,3$: %\begin{equation}\label{eq:a}
%{\mathbf a}_1=(1,0), \qquad {\mathbf a}_2=(1/2,\sqrt{3}/2),\qquad {\mathbf a}_3=(-1/2,\sqrt{3}/2)
%\end{equation}
%where we have set the lattice spacing $a$ to unity. 
\begin{equation}
\delta_{i}\equiv\theta_{{\mathbf{r}}+{\mathbf{a}}_{i}}-\theta_{{\mathbf{r}}}=\frac{1}{r}(\cos\theta,\sin\theta)\times{\mathbf{a}}_{i}.\label{eq:delta}
\end{equation}

We can then compute the correction to $\phi_{ab}$ to first order in $\delta_{i}$ using simple Taylor expansions.
The results of these calculations are represented graphically in Fig.~\ref{fig:trig}(c).
We observe that even in the presence of the variations $\delta_{i}$,
the Hamiltonian can be written as a $2\times2$ matrix in the same
basis as before: 
\[
H({\bf k})=H_{0}({\bf k})+V({\bf k}),
\]
where $V({\bf k})$ is the perturbation to the Hamiltonian due to the presence of the
vortex (which
depends on $\theta$ and $r$ and is valid to leading order in the
region shown in Fig.~\ref{fig:region}).

Moreover, we can represent the current operators $J_{x}({\bf k})=\frac{\partial H_{0}({\bf k})}{\partial k_{x}}$
and $J_{y}({\bf k})=\frac{\partial H_{0}({\bf k})}{\partial k_{y}}$,
as $2\times2$ matrices and write the tangential and radial current
operators as 
\[
J_{\theta}=-J_{x}\sin\theta+J_{y}\cos\theta,\qquad J_{r}=J_{x}\cos\theta+J_{y}\sin\theta.
\]
We are now ready to state the key result of this section. By explicitly
writing out $V({\bf k})$ (see Appendix~\ref{app:large_J} for details),
we find a relationship between the current operator and the perturbation
to the Hamiltonian: 
\begin{equation}
V({\bf k})=-J_{\theta}({\bf k})/2r+{\rm const.}\label{eq:key}
\end{equation}
This relationship was derived in a fixed (global) gauge and at an
operator level. Let us now consider a flux $\pi\equiv-\pi$ inserted
locally in the system. In the continuum limit, this local flux corresponds
to a tangential vector potential $A_{\theta}=\pi/2\pi r=1/2r$ along
a circle of radius $r$ and perimeter $2\pi r$ in some gauge. As
$A_{\theta}$ also couples to $J_{\theta}$, we find that the vortex effectively acts like a (fractional) flux $\pi$. Now, we know that if a flux $\pi$ in adiabatically inserted into an integer quantum Hall system with $\sigma_{xy}=e^{2}/h$, a fractional charge $1/2$ will be transported to the flux tube according to the Laughlin's
argument. Since the charge is a half the Hall conductance -- which is a topological invariant -- any small variation in the magnitude of the hopping amplitude, which we neglected earlier, can not change it.

One subtle issue with the above argument is that although we have a (gauge-dependent) operator relationship suggesting that the vortex effectively acts as a fractional flux, the gauge-invariant fluxes induced by the vortex are completely different from a localized flux $\pi$. 
In fact the vortex corresponds to an intricate pattern of
fluxes that only decays with the distance from the vortex core as $1/r$. The fractional charge does not correspond to a well-defined magnetic flux bound to the vortex. 
This means that the total flux through any closed loop around the vortex depends on the geometry of the loop and does not converge by increasing the loop size.
Despite this intricate flux pattern, we can perform a direct linear-response calculation for a dynamical process where the perturbation $V({\bf k})$ is turned on adiabatically, and as expected, we do obtain $q=1/2$. This direct calculation is presented in Appendix~\ref{app:large_J}.

\section{numerical results for charge and magnetization}
\label{sec:numeric}

\begin{figure}[ht]
\includegraphics[width=\columnwidth]{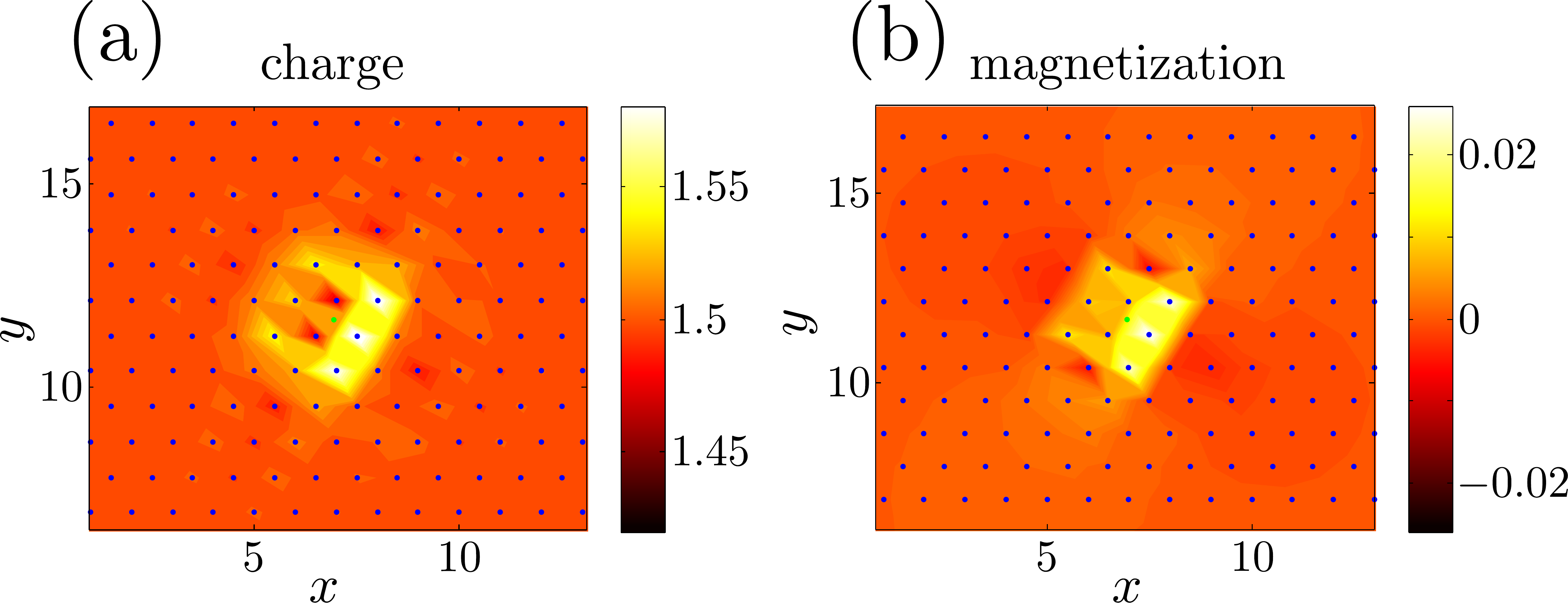}
\caption[]{Charge (a) and magnetization (b) density distributions around a vortex. The vortex is indicated by the green dot at the center. The system parameters are as follows: $J \to \infty$, $L =30$, $h=0$ at $3/4$ electronic filling.
} 
\label{FIG:2}
\end{figure}

\begin{figure}[ht]
\includegraphics[width=\columnwidth]{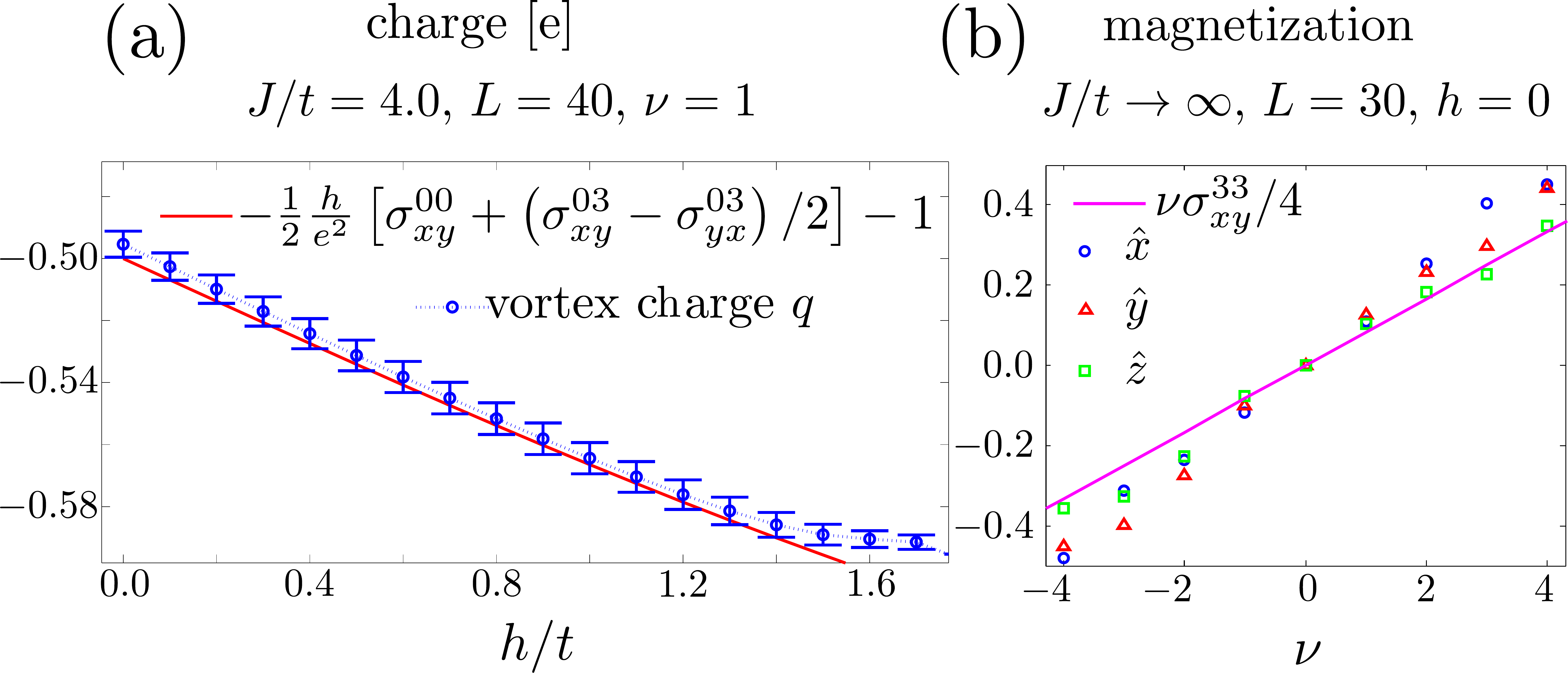}
\caption[]{
(a) Zeeman field dependence of the vortex charge. Blue circles were obtained by exact diagonalization. The red solid line is the expected result from a Laughlin adiabatic argument. (b) Net magnetization accumulated nearby the vortex for different windings and with rotations around the $z$ axes. Both results were obtained for $3/4$ electronic filling.}
\label{FIG:3}
\end{figure}

We now check the above results numerically.  We plot, in Fig.~\ref{FIG:2}, the charge and magnetization distribution in the vicinity of the vortex core for the triangular lattice model (the kagom\'e lattice gives similar distribution).
As expected, the charge localized in the core is half-odd-integer for odd winding and integer for even winding. 
The agreement between the vortex magnetization obtained numerically with what the Laughlin's argument would give is not perfect because the spin current is not conserved. 
Nevertheless, particularly for large $J$, the discrepancy is not too large and we verify that the vortex spin polarization approximately scales with the vortex winding, as shown in Fig.~\ref{FIG:3}(b).

We have also considered deviations from the fully symmetric assumptions. In particular, we added a Zeeman field  $h$ along $\hat{z}$ axis acting on electrons, i.e., $H\rightarrow H+h\sum_i    c_{i\alpha}^\dagger \sigma^z_{\alpha\beta} c_{i\beta}$. 
We verified that as long as the spectral gap does not close, the charge Hall conductivity does not change. However, in contrast to the fully symmetric case, new nonzero response functions emerge, namely $\sigma^{03}_{xy}, \,\sigma^{03}_{yx} \ne 0$, which correspond to the charge response to ${\cal A}_3$, the $\sigma_3$ component of the vector potential.
 In Fig.~\ref{FIG:3}(a), we plot the dependence of the vortex charge on $h$. The solid line is $q = -\sigma_{xy}/2 - (\sigma^{03}_{xy} - \sigma^{03}_{yx})/4-1$, which directly follows from the application of the Laughlin argument. Note that the charge is generally irrational and determined modulo an integer. The agreement is very good, all the way to the value of $h$ where the gap in the electronic spectrum closes. 
 
\begin{figure}[h]
\vspace{4mm}
 \includegraphics[width=\columnwidth]{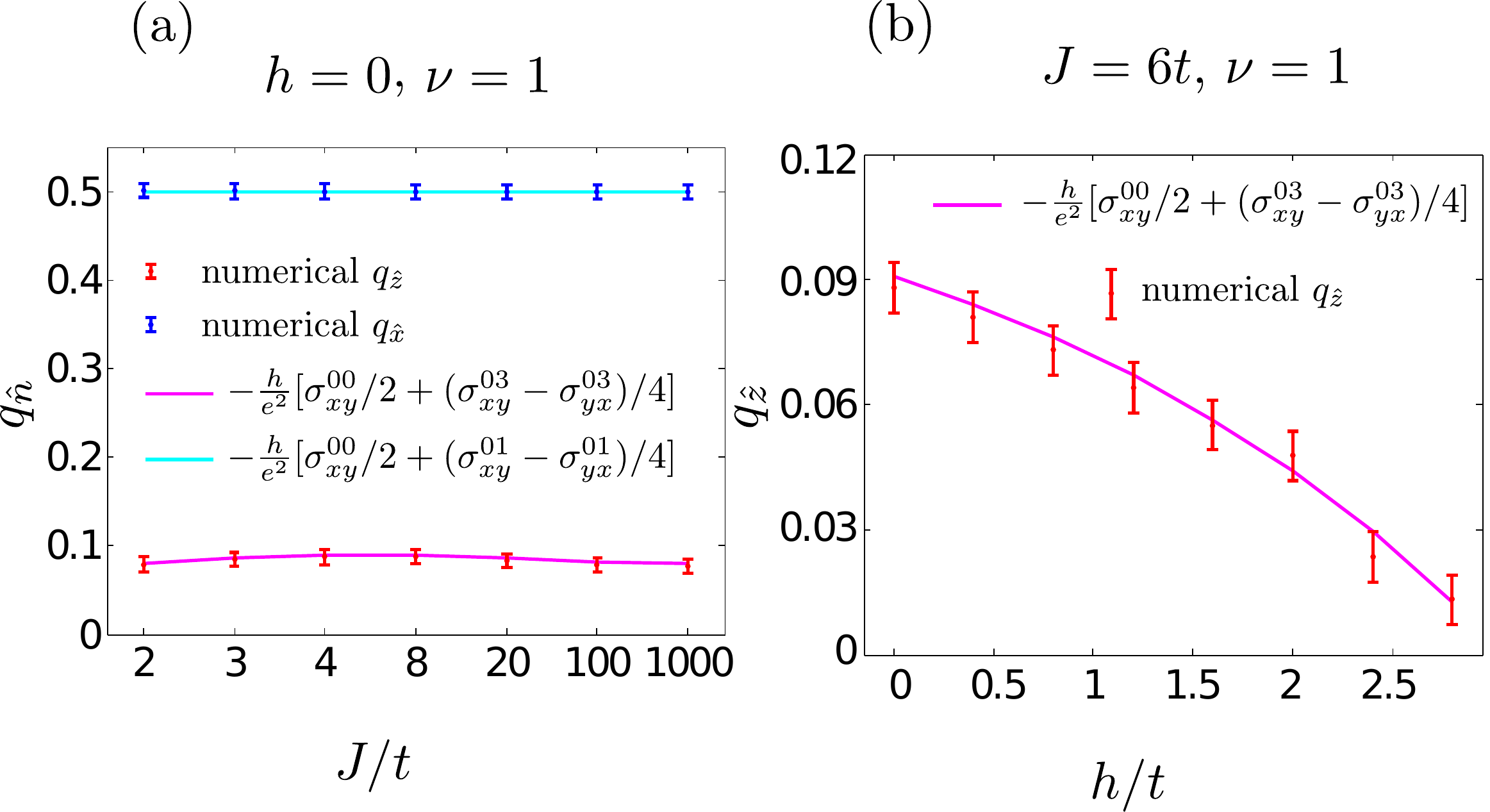}
\caption[]{ (a) Vortex charge for the kagom\'e lattice for different coupling strengths in the absence of a magnetic field. Red (blue) dots were obtained by exact diagonalization for a vortex with rotations around the $\hat{z}$ ($\hat{x}$) axis. The solid lines are the analytical predictions from the Laughlin's argument. As stated in the main text, the kagom\'e lattice has a net magnetization (in the $z$ direction), and stable vortices must have an axis of rotation pinned to the $z$ axis. Despite agreement with the Laughlin argument, the charge of the vortex with $\hat{x}$ axis is not stable, and will change with the relaxation of the vortex.
(b) Zeeman-field dependence of the vortex charge for the kagom\'e lattice. Red dots were obtained by exact diagonalization. The solid line is the result expected from Laughlin's adiabatic argument. The electronic filling is $5/6$ in both cases. 
 }
\label{fig:plot} 
\end{figure}

For the kagom\'e lattice case, we numerically computed the charge bound to a vortex with $\hat{n}=\hat{z}$ rotation axis, which is the only allowed axis for a bare texture with magnetization in the $z$ direction. 
Fig.~\ref{fig:plot}(a) shows the charge of a vortex in the kagom\'e lattice for several coupling strengths. The horizontal axis does not have a linear scale so both the variations at small $J$ and the saturation for large $J$ can be displayed.
The results computed from the Laughlin argument show very good agreement with the numerical ones. Interestingly, putting an \textit{ad hoc} vortex with the $\hat x$ axis of rotation also gives a charge consistent with the Laughlin argument (such \textit{ad hoc} vortex will not be stable and the charge will change once the vortex relaxes.)
Fig. \ref{fig:plot}(b) shows the charge $q_{\hat z}$ trapped by a vortex in the kagom\'e lattice as a function of the Zeeman field along $\hat{z}$ axis acting on electrons.
Just as in the triangular lattice case, we find that $\sigma^{00}_{xy}$ does not change as long as the spectral gap remains open. The off-diagonal responses $\sigma^{03}_{xy}$, on the other hand, change as a function of $h$, resulting in a continuously changing fractional charge.  
The results computed from the Laughlin's argument again show excellent agreement with the ones obtained by numerical diagonalization on finite lattices of size $24 \times 24$.

\section{exchange statistics}
\label{sec:berry}

We now turn to the exchange statistics of the vortices, considering the case of triangular lattice in zero Zeeman field for concreteness.
There are two distinct possibilities for a combination of two vortices: (1) the total charge of two vortices  is even, or (2) the total charge is odd.  The former case will be realized if the vortices are pulled apart from the uniform ``vacuum": since the initial state has total charge zero (relative to the uniform background), the state with two vortices will keep the same charge. Since the charge of an individual vortex is half-odd-integer, for large inter-vortex separation there are two energetically equivalent ground states that correspond to vortex charge configurations $(\,^1/_2, -\,^1/_2)$
and $(-\,^1/_2, \,^1/_2)$.  
This degeneracy can lead to non-Abelian exchange statistics, but unlike the Majorana states in the p-wave superconductor-like systems \cite{Read2000, Ivanov2001}, there is no topological protection. In other words, any local disorder can shift the bound state energy in a given vortex and lift the degeneracy.

The case (2) can be obtained, e.g., upon electron or hole doping of vortex bound states in the system with equally charged (half-odd-integer) vortices.
In this case, there is no ground-state degeneracy and exchanging two vortices can only give rise to an Abelian Berry phase. 
%The anyonic character of the vortices follows from a simple argument. 
%Consider two indistinguishable distant vortices with single winding and equal charge.
%Near each vortex the gauge potential is ${\bf A^\nu} =({1+ \sigma_3})\,\partial^\nu \phi/2$, with $\phi$ being the angular coordinate relative to the given vortex core. Therefore, each vortex has a non-Abelian gauge flux attached to it, equivalent to a $U(1)$ flux $2\pi$ in the spin-up channel. The number of up electrons occupying each vortex is $n_\uparrow = (q + 2m_z)/2$.  The statistical angle due to the flux attachment is the product of the charge and half of the flux, which for the idealized case of Eq. (\ref{eq:qsz}) is  $\theta = \pi (\sigma^{00}_{xy} + \sigma_{xy}^{33})/4$. Note that the deviations of $m_z$ from this idealized case are small for $J\rightarrow\infty$ (see $\nu=1$ data in Fig.~\ref{FIG:3}b). The numerical observation of the statistical angle, however, is complicated by the spin nonconservation.
%%%%%%%%%%%%%%
We argue that in this case (absence of degeneracies), the vortices have anyonic statistics
with a phase determined by the fractional charge and the magnetization
of the vortex. Since magnetization can depend on microscopic details
such as the location of the vortex core with respect to the lattice,
the statistical angle has path-dependent contributions. Despite these subtleties, we verify in this section, through extensive numerical calculations, that the statistical angle is indeed linearly related to the magnetization.

Consider vortices obtained by spin rotations along the $\hat{z}$ axis. As argued in the previous sections such vortices produce a non-Abelian flux
$
{\bf \Phi}=\oint{\bf A}d{\bf r}=({1+2m+\sigma_{3}})\pi$,
which is a $2\times2$ diagonal matrix coupled to spin-dependent density ${\bf n}=\left(\begin{array}{c}
n_{\uparrow}\\
n_{\downarrow}
\end{array}\right)$. Therefore we expect the total Berry phase accumulated upon adiabatically
taking one vortex all the way around the other to be equal to 
\[
{\rm tr}\left({\bf \Phi}{\bf n}\right)=(2m+1)\pi(n_{\uparrow}+n_{\downarrow})+\pi(n_{\uparrow}-n_{\downarrow})=(2m+1)\pi q+2\pi m_{z}.
\]
Noting that $q$ is half-odd integer, we obtain 
\begin{equation}
{\rm tr}\left({\bf \Phi}{\bf n}\right)=p\pi+\pi/2+2\pi m_{z},\label{eq:berry}
\end{equation}
where $p$ is an integer, which depends on microscopic details such
as the occupation of midgap states.

Let us comment that the above relationship is an exotic feature of
noncoplanar textures coupled to spinful electrons. In the more traditional
case, often discussed in the literature, fractional particles of charge
$1/2$ usually have statistical angle
${\pi}/{4}$. Indeed from our discussion in Sec.~\ref{sec:spinless},
we may naively expect such statistics. In the spinless model, we found
that one vortex effectively acts as a $U(1)$ flux $\pi$. Now taking
another vortex of charge $1/2$ around this vortex should result in
a Berry phase of ${\pi}/{2}$ and $\Theta=\pi/4$. What is
wrong with this argument? In the spinless case, we write a model and
construct wave functions in a basis labeled by lattice sites. Each
of the lattice sites, however, represent the projection of the electron
spin along the direction of the magnetic moment. As long as the vortices
are static, we do not need to worry about this additional information,
which is lost in the spinless model. However, if we have moving
vortices, working with a spinless model amounts to working in a time-dependent
basis.

To verify Eq.~\eqref{eq:berry}, we numerically compute the statistical angle
through exact diagonalization (for technical details see Appendix C). To find the statistical angle, we compute the Berry
phase $\phi_{{\rm bare}}$ acquired by taking a vortex along a closed
path, which does not enclose another vortex, and the Berry phase
$\phi_{{\rm vortex}}$ obtained by taking it around another vortex (on the same
path). The statistical
angle is then given by~\cite{Ryu2009}
\[
\theta=(\phi_{{\rm vortex}}-\phi_{{\rm bare}})/2.
\]

For efficiency, we have done our numerics in the limit of large $J$.
This allows us to diagonalize smaller matrices (by a factor of $2$),
but to compute the overlaps we need to put back in the information
regarding the local magnetic moments and expand our spinless wave
functions in a spinful basis: if we have amplitude $\varphi_{i}$
on site $i$ in the spinless wave function, and local moment $\vec{S}_{i}$
on that site, the amplitude in the spinful basis is simply $\varphi_{i}|\chi_{i}\rangle$,
where $|\chi_{i}\rangle$ is a spinor in the direction of $\vec{S}_{i}$.

We have performed our calculations
for three system sizes $L=40,50,60$ lattice spacing (see Fig.~\ref{fig:berry}
) with open boundary conditions (for brevity we only present data
for $L=60$ but similar conclusions can be drawn from smaller sizes). We have used path radii $r/L=0.22,0.24,0.26,0.28$,
and for each system size and path radius, we have used several numbers
of particles all inside the quarter-filling spectral gap with unoccupied
and occupied vortex-bound midgap states. We found that, as expected,
changing the particle number by filling edge modes does not affect
the Berry phase. Thus we present results for only two particle numbers:
one with all vortex bound states empty and one with a filled
bound state. We have performed our calculations with $512$ and $1024$
discretization point, and obtained good convergence. The results are shown in Table.~\ref{tab:vor}.
\begin{table}
\begin{centering}
\begin{small} %
\begin{tabular*}{1\linewidth}{@{\extracolsep{\fill}}@{\extracolsep{\fill}}@{\extracolsep{\fill}}cccccc}
\hline 
\hline
$L$  & $N$  & $r/L$  & $\phi_{{\rm bare}}$  & $\phi_{\rm vortex}$  & $\phi_{\rm vortex}-\phi_{{\rm bare}}$ \bigstrut \tabularnewline
\hline 
60  & 1796  & 0.22  & 3.81  & 2.85  & 5.32\bigstrut \tabularnewline
60  & 1796  & 0.24  & 0.70  & 5.65  & 4.95\bigstrut \tabularnewline
60  & 1796  & 0.26  & 5.96  & 4.68  & 5.00\bigstrut \tabularnewline
60  & 1796  & 0.28  & 2.08  & 1.08  & 5.29\bigstrut \tabularnewline
\hline 
60  & 1816  & 0.22  & 3.58  & 5.98  & 2.40\bigstrut \tabularnewline
60  & 1816  & 0.24  & 4.44  & 0.32  & 2.17\bigstrut \tabularnewline
60  & 1816  & 0.26  & 0.76  & 2.81  & 2.05\bigstrut \tabularnewline
60  & 1816  & 0.28  & 0.32  & 2.58  & 2.26\bigstrut \tabularnewline
\hline 
\hline
\end{tabular*}\end{small} 
\par\end{centering}

\caption{\label{tab:vor}Berry phases for taking one vortex around another. The
results are not completely path independent but show reasonable stability.}
\end{table}

The results above do show path-dependent fluctuations
of around $10-15\%$. However, there is good stability for different
system sizes. Since we expect the Berry phase to depend on the magnetization
(that is not a sharp quantum number and can fluctuate), this is not
surprising. Moreover, our results suffer from finite-size effects:
the charge profile of the two vortices may overlap with each other
and the charges accumulated at the edge of the finite system.

As a benchmark for our method, we also performed calculations for
the Berry phase accumulated by taking the vortex around a local flux
$\pi$ inserted in one triangular plaquette. The results are shown in Table.~\ref{tab:pi}. In this case, for a vortex
of charge $q$, we expect a Berry phase of $\phi_{\pi{\rm -flux}}-\phi_{{\rm bare}}=(2n+1)\pi q$
where $n$ is an integer. With (without) an occupied vortex bound
state the charge of the vortex was $1/2$ ($-1/2$). Interestingly,
for both cases we found a Berry phase close to $+\pi/2$, which is
consistent with the above expression for $n=0$ and $n=-1$ respectively
(flux $\pi$ is equivalent to flux $-\pi$). The benchmark above shows that while our numerical method is capable
of reproducing established results, there is error of order a few
percent in the finite-size numerical lattice calculation. 
\begin{table}
\begin{centering}
\begin{small} %
\begin{tabular*}{1\linewidth}{@{\extracolsep{\fill}}@{\extracolsep{\fill}}@{\extracolsep{\fill}}cccccc}
\hline 
\hline
$L$  & $N$  & $r/L$  & $\phi_{{\rm bare}}$  & $\phi_{\pi{\rm -flux}}$  & $\phi_{\pi{\rm -flux}}-\phi_{{\rm bare}}$ \bigstrut \tabularnewline
\hline 
60  & 1796  & 0.22  & 3.81  & 5.34  & 1.53\bigstrut \tabularnewline
60  & 1796  & 0.24  & 0.70  & 2.23  & 1.53\bigstrut \tabularnewline
60  & 1796  & 0.26  & 5.96  & 1.20  & 1.53\bigstrut \tabularnewline
60  & 1796  & 0.28  & 2.08  & 3.61  & 1.53\bigstrut \tabularnewline
\hline 
60  & 1816  & 0.22  & 3.58  & 5.20  & 1.62\bigstrut \tabularnewline
60  & 1816  & 0.24  & 4.44  & 6.05  & 1.62\bigstrut \tabularnewline
60  & 1816  & 0.26  & 0.76  & 2.38  & 1.62\bigstrut \tabularnewline
60  & 1816  & 0.28  & 0.32  & 1.94  & 1.62\bigstrut \tabularnewline
\hline 
\hline 
\end{tabular*}\end{small} 
\par\end{centering}

\caption{\label{tab:pi}Berry phases for taking a vortex around a local flux
$\pi$. The results show good agreement with the theoretical
prediction of $\phi_{\pi{\rm -flux}}-\phi_{{\rm bare}}=\pi/2$. For
each system size, we have two different values of the number of particles
$N$. We have $q=1/2$ ($q=-1/2$ for the larger (smaller) of the
two values of $N$. }
\end{table}

\begin{figure}[ht]
\includegraphics[width=0.9\columnwidth]{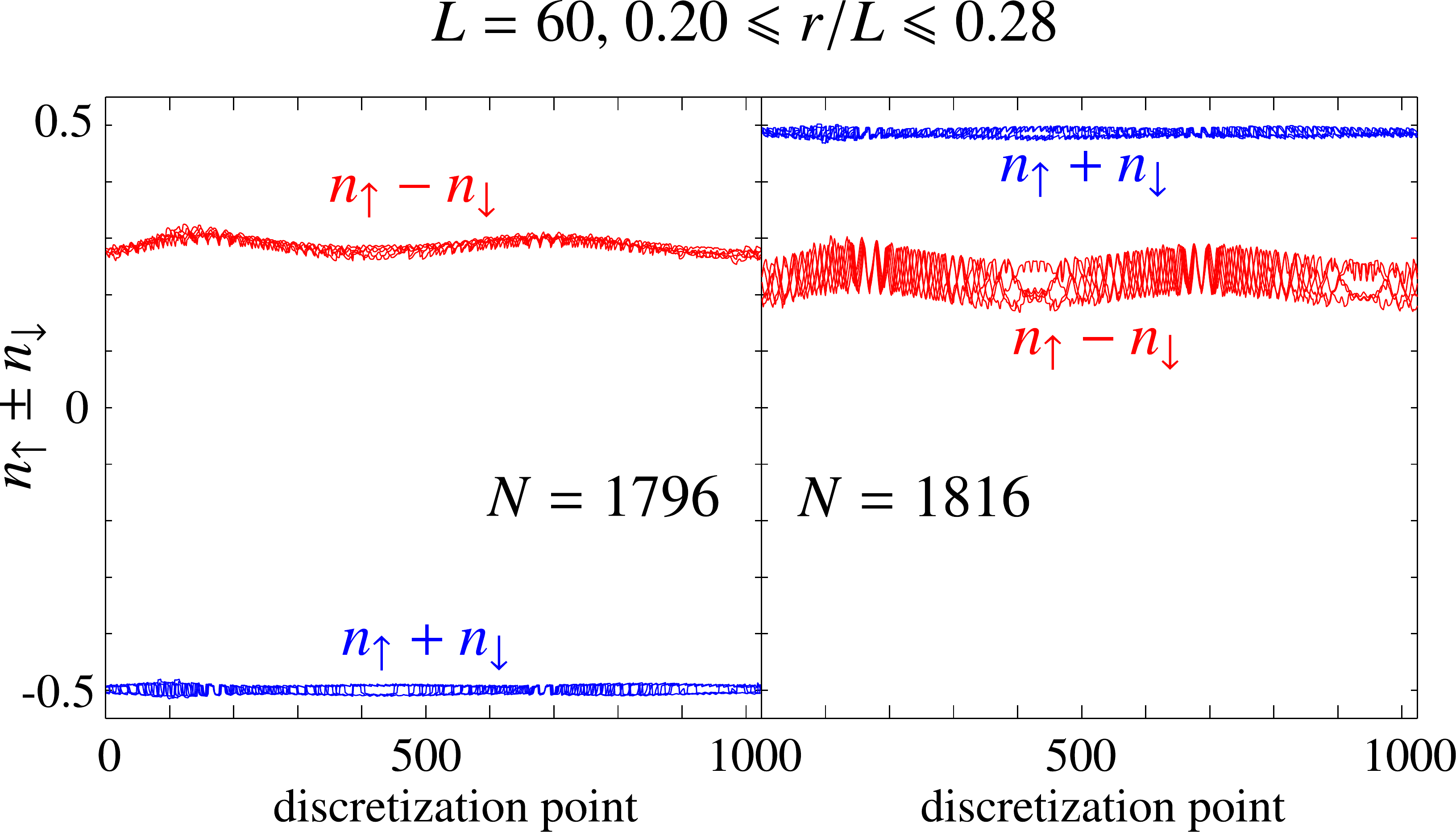} \caption[]{The measures charge and magnetization of the vortex as it moves around
another in a system of $L=60$ for different path radii $r/L=0.20, 0.22, 0.24, 0.26, 0.28$.}
\label{fig:mag} 
\end{figure}

We now present
direct measurements of $n_{\uparrow}+n_{\downarrow}$ and $n_{\uparrow}-n_{\downarrow}$
in a box of size $3.5\times3.5$ centered at the core of the moving
vortex. The results are shown in Fig.~\ref{fig:mag}. We see that
the measured charge exhibits some fluctuations (a few percent) as a function of the
position of the vortex core (on a single path and between different
paths) due to the finite-size effects. The fluctuations of the magnetization are larger due to the
nonconservation of spin. In particular when a vortex bound state is
occupied, the local physics can strongly affect the magnetization of
a vortex. Despite all these issues, the Berry phases we measured directly
show good agreement with Eq.~\eqref{eq:berry}. In particular,
if we average the fluctuating magnetization over the full cyclic path, the result
is independent of the path. We can now compute the expected Berry
phases from Eq.~\eqref{eq:berry}. The results are shown below and
their agreement with Table \ref{tab:vor} strongly supports the correctness
of Eq.~\eqref{eq:berry}.

\begin{center}
\begin{small} %
\begin{tabular*}{1\linewidth}{@{\extracolsep{\fill}}@{\extracolsep{\fill}}@{\extracolsep{\fill}}ccccc}
\hline 
$L$  & $N$  & $q$  & $\overline{n_{\uparrow}-n_{\downarrow}}$  & $\phi=\pi(q+\overline{n_{\uparrow}-n_{\downarrow}})$\bigstrut \tabularnewline
\hline 
60  & 1796  & -1/2  & 0.28  & 5.59\bigstrut \tabularnewline
60  & 1816  & +1/2  & 0.23  & 2.29\bigstrut\tabularnewline
\hline 
\end{tabular*}\end{small} 
\par\end{center}

\section{conclusion}

In summary, we have shown that integer quantum Hall systems, which emerge from the interplay of itinerant electrons and noncoplanar magnetic ordering, generically support topologically stable excitations with fractional charge and anyonic statistics. We showed through extensive numerical calculations that the statistical angle of these anyons has a simple relationship to their charge and magnetization.

The energetics of $Z_2$ vortices is similar to the usual $Z$ vortices for $SO(2)$ order parameter~\cite{Kawamura1984}: i.e., the energy of an isolated vortex scales logarithmically with the system size. Nevertheless, pairs of log-confined vortex pairs will appear due to thermal fluctuations at finite temperatures. 
In addition, inclusion of quantum spin dynamics may lead to an intriguing possibility that the {\em quantum} fluctuations transform the noncoplanar ordered state into a chiral spin liquid~\cite{Kalmeyer1987, Wen1989, Wiegmann1992} at zero temperature. There, the fractionally charged $Z_2$ vortices discussed above may turn into deconfined point-like excitations.

 Promising candidates materials which may exhibit this physics could be the systems of Na$_x$CoO$_2$ type, which near $x = 0.5$ are known to have a noncollinear order, as well as anomalously large Hall response \cite{Foo2004}. The fractional charge predicted in this work may be accessible through direct imaging of the local charge profile, as shown in Fig.~\ref{FIG:2}, e.g., by scanning force microscopy. Also anyonic exchange statistics may have unusual consequences in real materials. Perhaps the most intriguing among them is the possibility of the Anyonic superconductivity \cite{Laughlin1988, Wilczek1990, Wiegmann1992}.
When a system is doped away from the chiral Mott insulating state, it may energetically prefer to  accommodate the carriers by creating vortices with intragap states.  As we have just argued, such occupied vortex states are anyons, which, at finite density and low enough temperature, may go into a superconducting state \cite{Wilczek1990}.

%%%%%%%%%%%%%%

\section{acknowledgements} 
We thank  C. Batista, C. Chamon, C.-Y. Hou, D. Ivanov, A. Morpurgo, D. Podolsky, S. Ryu, S. Sachdev and L. Santos for helpful discussions.  This work was carried out under the auspices of the National Nuclear Security Administration of the U.S. Department of Energy at Los Alamos National Laboratory under Contract No. DE-AC52-06NA25396 and supported by the LANL/LDRD Program. R. Muniz also thanks CNPq (Brazil) for financial support.

\appendix
%dummy comment inserted by tex2lyx to ensure that this paragraph is not empty

\section{Details of the $J \to \infty$ microscopic derivation}
\label{app:large_J}

\subsection{Derivation of Eq.~\eqref{eq:key}}
Here we show the details of
the derivation of Eq.~\eqref{eq:key} by explicitly writing out the
terms appearing in two sides of this relationship. First, Eq.~\eqref{eq:delta}
for the three lattice vectors ${\bf a}_{i}$ gives 
\begin{eqnarray}\label{eq:delta2}
\delta_{1} & \equiv & \delta_{{\mathbf{r}}+{\mathbf{a}}_{1}}-\delta_{{\mathbf{r}}}=-{\sin\theta/r},\nonumber \\
\delta_{2} & \equiv & \delta_{{\mathbf{r}}+{\mathbf{a}}_{1}}-\delta_{{\mathbf{r}}}=\left(\frac{\sqrt{3}}{2}\cos\theta-\frac{1}{2}\sin\theta\right)/r,\nonumber \\
\delta_{3} & \equiv & \delta_{{\mathbf{r}}+{\mathbf{a}}_{1}}-\delta_{{\mathbf{r}}}=\left(\frac{\sqrt{3}}{2}\cos\theta+\frac{1}{2}\sin\theta\right)/r.
\end{eqnarray}
The Hamiltonian in the presence of the above $\delta_{1}$ can then
be written as

\begin{widetext} 
\begin{equation}
H({\mathbf{k}})=-2\left(\begin{array}{cc}
\cos\left({\mathbf{k}}\cdot{\mathbf{a}}_{2}+\frac{\sqrt{3}-1}{2}\delta_{2}\right) & e^{i\pi/4}\cos\left({\mathbf{k}}\cdot{\mathbf{a}}_{1}-\delta_{1}/2\right)+e^{-i\pi/4}\cos\left({\mathbf{k}}\cdot{\mathbf{a}}_{3}-\delta_{3}/2\right)\\
e^{-i\pi/4}\cos\left({\mathbf{k}}\cdot{\mathbf{a}}_{1}-\delta_{1}/2\right)+e^{i\pi/4}\cos\left({\mathbf{k}}\cdot{\mathbf{a}}_{3}-\delta_{3}/2\right) & -\cos\left({\mathbf{k}}\cdot{\mathbf{a}}_{2}-\frac{\sqrt{3}+1}{2}\delta_{2}\right)
\end{array}\right),
\end{equation}
which upon expanding to linear order in $\delta_{j}$ gives $H({\mathbf{k}})=H_{0}({\mathbf{k}})+V({\mathbf{k}})$
with 
\begin{equation}
H_{0}({\mathbf{k}})=-2\left(\begin{array}{cc}
\cos\left({\mathbf{k}}\cdot{\mathbf{a}}_{2}\right) & e^{i\pi/4}\cos\left({\mathbf{k}}\cdot{\mathbf{a}}_{1}\right)+e^{-i\pi/4}\cos\left({\mathbf{k}}\cdot{\mathbf{a}}_{3}\right)\\
e^{-i\pi/4}\cos\left({\mathbf{k}}\cdot{\mathbf{a}}_{1}\right)+e^{i\pi/4}\cos\left({\mathbf{k}}\cdot{\mathbf{a}}_{3}\right) & -\cos\left({\mathbf{k}}\cdot{\mathbf{a}}_{2}\right)
\end{array}\right),
\end{equation}
and 
\begin{equation}
V({\mathbf{k}})=\left(\begin{array}{cc}
\left(\sqrt{3}-1\right)\sin\left({\mathbf{k}}\cdot{\mathbf{a}}_{2}\right)\delta_{2} & -e^{i\pi/4}\sin\left({\mathbf{k}}\cdot{\mathbf{a}}_{1}\right)\delta_{1}-e^{-i\pi/4}\sin\left({\mathbf{k}}\cdot{\mathbf{a}}_{3}\right)\delta_{3}\\
-e^{-i\pi/4}\sin\left({\mathbf{k}}\cdot{\mathbf{a}}_{1}\right)\delta_{1}-e^{i\pi/4}\sin\left({\mathbf{k}}\cdot{\mathbf{a}}_{3}\right)\delta_{3} & \left(\sqrt{3}+1\right)\sin\left({\mathbf{k}}\cdot{\mathbf{a}}_{2}\right)\delta_{2}
\end{array}\right).\label{eq:V}
\end{equation}
The current operators can be obtained by differentiating $H_{0}({\bf k})$ with respect to ${\bf k}$, and are given by 
\begin{equation}
\begin{split} & J_{x}({\mathbf{k}})=\left(\begin{array}{cc}
\sin\left({\mathbf{k}}\cdot{\mathbf{a}}_{2}\right) & 2\: e^{i\pi/4}\sin\left({\mathbf{k}}\cdot{\mathbf{a}}_{1}\right)-e^{-i\pi/4}\sin\left({\mathbf{k}}\cdot{\mathbf{a}}_{3}\right)\\
2\: e^{-i\pi/4}\sin\left({\mathbf{k}}\cdot{\mathbf{a}}_{1}\right)-e^{i\pi/4}\sin\left({\mathbf{k}}\cdot{\mathbf{a}}_{3}\right) & -\sin\left({\mathbf{k}}\cdot{\mathbf{a}}_{2}\right)
\end{array}\right),\\
 & J_{y}({\mathbf{k}})=\left(\begin{array}{cc}
\sqrt{3}\:\sin\left({\mathbf{k}}\cdot{\mathbf{a}}_{2}\right) & \sqrt{3}\: e^{-i\pi/4}\sin\left({\mathbf{k}}\cdot{\mathbf{a}}_{3}\right)\\
\sqrt{3}\: e^{i\pi/4}\sin\left({\mathbf{k}}\cdot{\mathbf{a}}_{3}\right) & -\sqrt{3}\:\sin\left({\mathbf{k}}\cdot{\mathbf{a}}_{2}\right)
\end{array}\right).
\end{split}
\end{equation}
Inserting Eq.~(\ref{eq:delta2}) into Eq.~(\ref{eq:V}) and comparing
with $J_{x}({\mathbf{k}})$ and $J_{y}({\mathbf{k}})$ above leads
to Eq.~\eqref{eq:key}:
\begin{equation}
V({\mathbf{k}})=\sqrt{3}\sin\left({\mathbf{k}}\cdot{\mathbf{a}}_{2}\right)\delta_{2}\times\openone+\frac{\sin\theta}{2\: r}\: J_{x}({\mathbf{k}})-\frac{\cos\theta}{2\: r}\: J_{y}({\mathbf{k}}).
\end{equation}
\end{widetext}

\subsection{Asymptotic flux pattern}
As mentioned before, despite the relationship~\eqref{eq:key}, the flux pattern is different than a localized flux $\pi$ at the center of the vortex. Here, we explicitly show this intricate pattern. Far away from the vortex, the flux pattern has translation invariance
(to leading order) within local regions defined in Fig.~\ref{fig:region} and
can be easily calculated for a vortex with any axis of rotation $\vec{l}$
(so far we have only focused on $\vec{l}=\hat{z}$). Consider a triangular
plaquette with three magnetic moments $\vec{S}_{a,b,c,}$. If the
effective flux through the plaquette is $\phi$, we have the
solid angle formula
\begin{equation}
\cot(\phi)=\frac{1+\vec{S}_{a}\cdot\vec{S}_{b}+\vec{S}_{b}\cdot\vec{S}_{c}+\vec{S}_{c}\cdot\vec{S}_{a}}{|\vec{S}_{a}\:\vec{S}_{b}\:\vec{S}_{c}|},\label{eq:solid}
\end{equation}
where $|\vec{S}_{a}\:\vec{S}_{b}\:\vec{S}_{c}|$ indicates the determinant
of a matrix $A_{ij}=S_{i}(j)$ (the $j$th component of $S_i$). It is easy to see that the numerator
vanishes for any triangle in the unperturbed magnetic structure of
Fig.~\ref{fig:trig}(a), as it should for  $\pi/2$ flux per triangular plaquette. %\begin{figure}[ht]
% \includegraphics[width=3.5cm]{plaq}
%\caption[]{The labelling convention for up and down triangular plaquettes. In the magnetic texture of Fig.~\ref{fig:lattice}, there are four types of up and down triangles depending on what the moments $\vec{S}_{a,b,c,}$ are.}
%\label{fig:plaq} 
%\end{figure}
Now if a moment $\vec{S}_{i}$ is rotated by an angle $\theta_{i}$
around $\vec{l}$, we can show after some algebra that $\vec{S}_{a}\cdot\vec{S}_{b}\rightarrow\vec{S}_{a}\cdot\vec{S}_{b}\cos\delta_{ab}+\vec{l}\cdot\left(\vec{S}_{a}\times\vec{S}_{b}\right)\sin\delta_{ab}+\:\left(\vec{l}\cdot\vec{S}_{a}\right)\left(\vec{l}\cdot\vec{S}_{b}\right)\left(1-\cos\delta_{ab}\right)$.
For small $\delta_{ab}=\theta_{a}-\theta_{b}$ {[}see Eq.~\eqref{eq:delta}{]},
the leading correction comes from the second term. The leading
correction to the flux then follows from the solid-angle  formula {[}see
Eq.~\eqref{eq:solid}{]}, and the results are shown in Table~\ref{tab:flux}.
\begin{table}
\centering{}%
\begin{tabular*}{1\linewidth}{@{\extracolsep{\fill}}@{\extracolsep{\fill}}@{\extracolsep{\fill}}ccc}
\hline 
type  & $(a,b,c)$  & $\delta\phi$ \tabularnewline
\hline 
\hline 
 &  & \tabularnewline
\includegraphics[width=0.7cm]{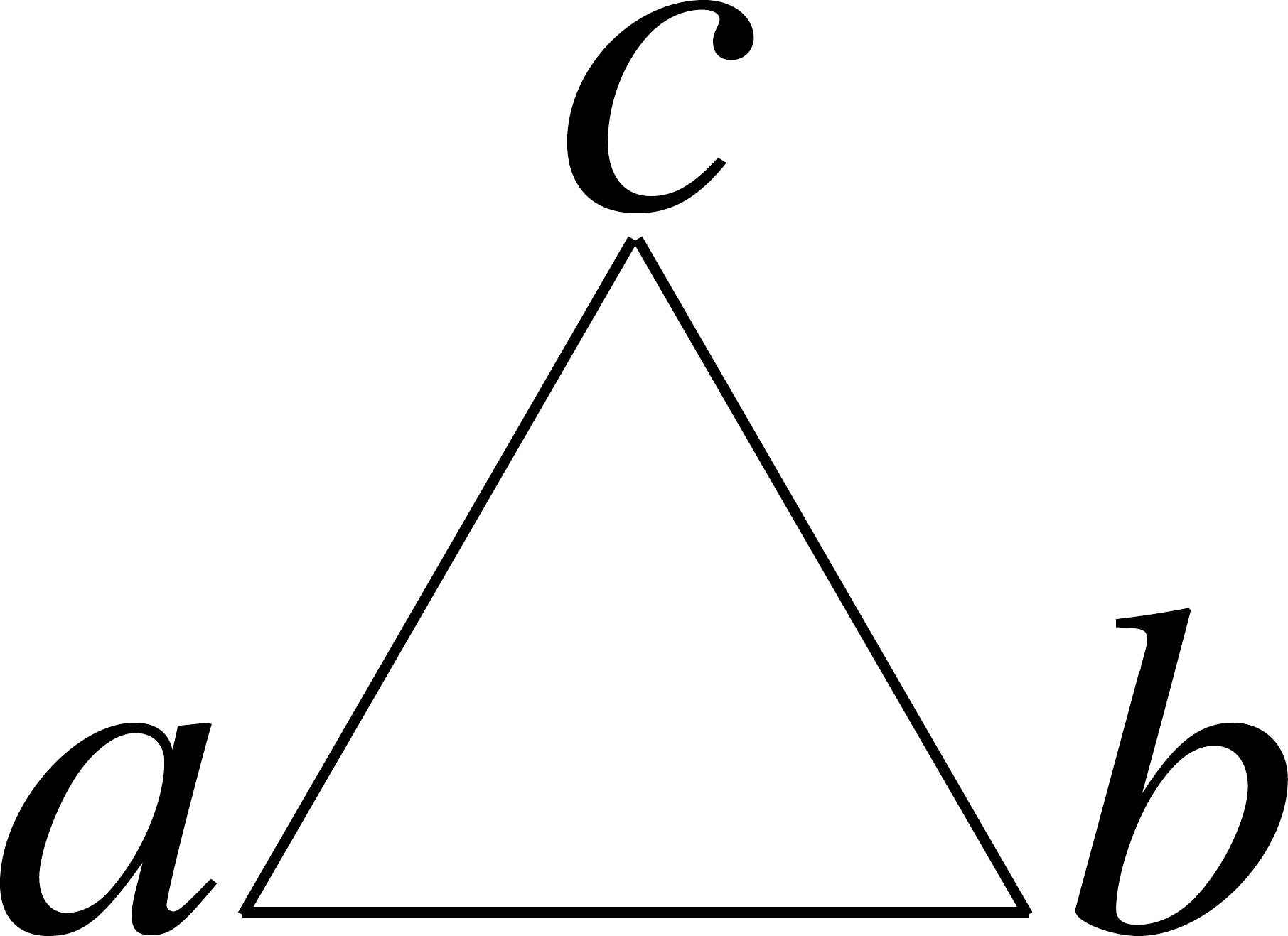}  & $(1,2,4)$  & $-{\sqrt{3}}\:\vec{l}\cdot\left(\theta_{a}-\theta_{b},\:\theta_{c}-\theta_{b},\:\theta_{c}-\theta_{a}\right)/2$
\bigstrut \tabularnewline
\includegraphics[width=0.7cm]{up}  & $(2,1,3)$  & $-{\sqrt{3}}\:\vec{l}\cdot\left(\theta_{a}-\theta_{b},\:\theta_{b}-\theta_{c},\:\theta_{a}-\theta_{c}\right)/2$
\bigstrut \tabularnewline
\includegraphics[width=0.7cm]{up}  & $(4,3,1)$  & $-{\sqrt{3}}\:\vec{l}\cdot\left(\theta_{a}-\theta_{b},\:\theta_{b}-\theta_{c},\:\theta_{c}-\theta_{a}\right)/2$
\bigstrut \tabularnewline
\includegraphics[width=0.7cm]{up}  & $(3,4,2)$  & $-{\sqrt{3}}\:\vec{l}\cdot\left(\theta_{b}-\theta_{a},\:\theta_{c}-\theta_{b},\:\theta_{a}-\theta_{c}\right)/2$
\bigstrut \tabularnewline
\hline 
\includegraphics[width=0.7cm]{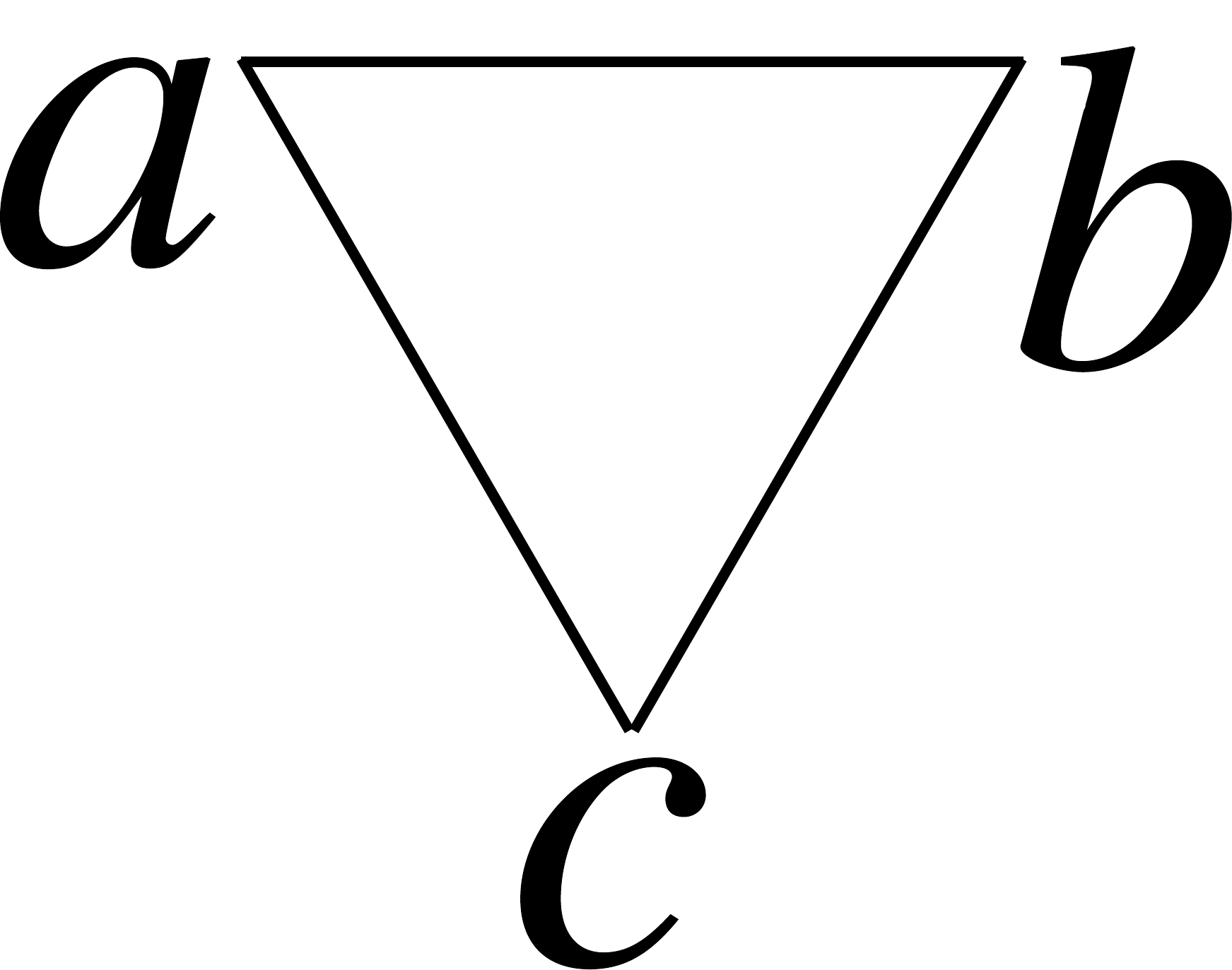}  & $(4,3,2)$  & $+{\sqrt{3}}\:\vec{l}\cdot\left(\theta_{b}-\theta_{a},\:\theta_{c}-\theta_{a},\:\theta_{b}-\theta_{c}\right)/2$
\bigstrut \tabularnewline
\includegraphics[width=0.7cm]{down}  & $(3,4,1)$  & $+{\sqrt{3}}\:\vec{l}\cdot\left(\theta_{a}-\theta_{b},\:\theta_{a}-\theta_{c},\:\theta_{c}-\theta_{b}\right)/2$\bigstrut \tabularnewline
\includegraphics[width=0.7cm]{down}  & $(1,2,3)$  & $+{\sqrt{3}}\:\vec{l}\cdot\left(\theta_{b}-\theta_{a},\:\theta_{a}-\theta_{c},\:\theta_{b}-\theta_{c}\right)/2$
\bigstrut \tabularnewline
\includegraphics[width=0.7cm]{down}  & $(2,1,4)$  & $+{\sqrt{3}}\:\vec{l}\cdot\left(\theta_{a}-\theta_{b},\:\theta_{c}-\theta_{a},\:\theta_{c}-\theta_{b}\right)/2$
\bigstrut \tabularnewline
\hline 
\end{tabular*}\caption{ \label{tab:flux} Asymptotic additional fluxes in triangular plaquettes
due to the presence of a vortex. }
\end{table}

\subsection{Direct linear response}

Given the intricate flux pattern, it is helpful to give a more microscopic derivation
of the fractional charge using Eq.~\eqref{eq:key} and the Laughlin's
argument. Basically, we want to compute the expectation value of the
current flowing toward the vortex core (i.e., $-\langle J_{r}\rangle$)
in linear response. Instead of adiabatically inserting a local flux,
however, in this case, we insert an intricate pattern of fluxes globally.
In the region of Fig.~\ref{fig:region}, inserting this flux pattern
corresponds to adiabatically turning on the perturbation $V({\mathbf{k}})$
of Eq.~(\ref{eq:key}) from zero to its final value. Let us consider
a linear in time protocol with total time $T$ as follows 
\begin{equation}
{V}(t)=\sum_{\mathbf{k}}\Psi_{\mathbf{k}}^{\dagger}{V}({\mathbf{k}},t)\:\Psi_{\mathbf{k}},\qquad{V}({\mathbf{k}},t)=\frac{t}{T}V({\mathbf{k}})\label{eq:Vt}
\end{equation}
It is worth mentioning that it is common in the literature to add a term of the form $e^{\epsilon t} H'$ with $\epsilon \rightarrow 0^+$ to the Hamiltonian in order to model adiabatically turning on a final perturbation $H'$ from $t=-\infty$ to any finite $t$. Here, we use the above linear protocol, which is convenient for the case of quantum Hall response.

Note that, to invoke adiabaticity, we need the spectral gap to remain
open during this process. We have checked numerically that the gap
remains open if all the additional fluxes (due to the presence of
the vortex) in each triangular plaquette is turned on linearly in
time. The linear-response expression for the expectation value
of an operator (in the present case $J_{r}$) at time $t$ is given by 
\begin{equation}
\langle J_{r}(t)\rangle=i\int_{0}^{t}dt^{\prime}\langle0|\left[\hat{V}(t^{\prime}),\hat{J}_{r}(t)\right]|0\rangle,\label{eq:LR}
\end{equation}
where $|0\rangle$ is the initial state (in this case the ground state
of $H_{0}$ in the absence of a vortex) and the ``hat'' notation
represents an operator in the interaction picture with the bare Hamiltonian
$H_{0}$, e.g., $\hat{V}(t^{\prime})=e^{iH_{0}t^{\prime}}V(t^{\prime})e^{-iH_{0}t^{\prime}}$,
with $V(t^{\prime})$ defined in Eq.~(\ref{eq:Vt}). 

To proceed, we diagonalize $H_{0}$ as follows: $H_{0}=\sum_{{\mathbf{k}},a=1,2}\varepsilon_{\mathbf{k}}^{a}\gamma_{a{\mathbf{k}}}^{\dagger}\gamma_{a{\mathbf{k}}}$,
where $\varepsilon_{\mathbf{k}}^{1,2}=\pm2\:\sqrt{\cos^{2}{\mathbf{k}}\cdot{\mathbf{a}}_{1}+\cos^{2}{\mathbf{k}}\cdot{\mathbf{a}}_{2}+\cos^{2}{\mathbf{k}}\cdot{\mathbf{a}}_{3}}$
are the eigenvalues of $H_{0}({\mathbf{k}})$ with corresponding orthonormal
eigenvectors $|1_{\mathbf{k}}\rangle$ and $|2_{\mathbf{k}}\rangle$.
% A unitary matrix $U_{\mathbf k}$ which diagonalizes the Hamiltonian $H_0$ is constructed from these eigenvectors and defining
%\[
%\Gamma_{\mathbf k}=U^\dagger_{\mathbf k}\Psi_{\mathbf k}, \qquad \Gamma^\dagger_{\mathbf k}=\left (\gamma^\dagger_{1{\mathbf k}}\,\gamma^\dagger_{2{\mathbf k}}\right),
%\]  
%and using $\gamma_{m{\mathbf k}}(t)\equiv e^{i H_0 t}\gamma_{m{\mathbf k}}e^{-i H_0 t}=e^{-i\varepsilon^{m}_{\mathbf k}t}\gamma_{m{\mathbf k}}$, we can write Eq.~(\ref{eq:LR}) as follows
We can now write Eq.~(\ref{eq:LR}) as 
\begin{equation}
\begin{split} & \langle J_{r}(t)\rangle=\sum_{{\mathbf{k}}{\mathbf{k}}^{\prime}\: mn\: m^{\prime}n^{\prime}}i\int_{0}^{t}dt^{\prime}\frac{t^{\prime}}{T}\times\\
 & \langle0|\Big[\gamma_{m{\mathbf{k}}}^{\dagger}e^{i\varepsilon_{\mathbf{k}}^{m}t}\: V^{mn}({\mathbf{k}})\:\gamma_{n{\mathbf{k}}}e^{-i\varepsilon_{\mathbf{k}}^{n}t},\gamma_{{m^{\prime}}{{\mathbf{k}}^{\prime}}}^{\dagger}e^{i\varepsilon_{{\mathbf{k}}^{\prime}}^{m^{\prime}}t}\: J_{r}^{{m^{\prime}}{n^{\prime}}}({\mathbf{k}})\:\gamma_{{n^{\prime}}{{\mathbf{k}}^{\prime}}}e^{-i\varepsilon_{{\mathbf{k}}^{\prime}}^{n^{\prime}}t}\Big]|0\rangle,
\end{split}
\end{equation}
where $O^{mn}({\mathbf{k}})$ indicates $\langle m_{\mathbf{k}}|O({\mathbf{k}})|n_{\mathbf{k}}\rangle$.
Summation over $m,n,m',n'$ gives 16 types of commutators. It turns
out, however, that only the following two types of commutators have
a nonvanishing contribution (i.e., reduce to $\gamma^{\dagger}\gamma$):
\begin{eqnarray*}
\left[\gamma_{1{\mathbf{k}}}^{\dagger}\gamma_{2{\mathbf{k}}},\gamma_{2{{\mathbf{k}}^{\prime}}}^{\dagger}\gamma_{1{{\mathbf{k}}^{\prime}}}\right]=\delta_{{\mathbf{k}}{{\mathbf{k}}^{\prime}}}\left(\gamma_{1{\mathbf{k}}}^{\dagger}\gamma_{1{\mathbf{k}}}-\gamma_{2{\mathbf{k}}}^{\dagger}\gamma_{2{\mathbf{k}}}\right),
\end{eqnarray*}
and a similar commutator with $1$ and $2$ indices switched. Defining
$n_{\mathbf{k}}^{m}=\langle0|\gamma_{m{\mathbf{k}}}^{\dagger}\gamma_{m{\mathbf{k}}}|0\rangle$,
this leads to 
\begin{eqnarray*}
\langle J_{r}(t)\rangle=i\sum_{{\mathbf{k}}mn}\int_{0}^{t}dt^{\prime}\frac{t^{\prime}}{T}e^{i(\varepsilon_{\mathbf{k}}^{m}-\varepsilon_{\mathbf{k}}^{n})(t^{\prime}-t)}\left(n_{\mathbf{k}}^{m}-n_{\mathbf{k}}^{n}\right)V^{mn}({\mathbf{k}})J_{r}^{nm}({\mathbf{k}}).
\end{eqnarray*}
%which upon integration by parts yields
%$\langle J_r (t)\rangle={i\over T}\sum_{{\mathbf k}mn}
%{V^{mn} ({\mathbf k})J_{r}^{nm} ({\mathbf k})\over \left(\varepsilon^{m}_{\mathbf k}-\varepsilon^{n}_{\mathbf k}\right)^2}
%\left( n^m_{\mathbf k}-n^n_{\mathbf k}\right)$.

The integral over $t'$ can be simply done by integration by parts.
We then need to integrate the resulting expression, which is independent
of time, over time from $0$ to $T$ and over a circle of radius $r$
for $0<\theta<2\pi$, which, putting back the factors of $e$ and
$\hbar$ gives 
\begin{equation}
q=rT\int_{0}^{2\pi}d\theta\langle J_{r}(t)\rangle=\frac{1}{2}\frac{h}{e}\sigma_{xy}=e/2,
\end{equation}
where the quantized Hall conductance $\sigma_{xy}$ is given by 
\begin{equation}\label{eq:cond}
\sigma_{xy}=\frac{e^{2}}{h}\frac{i}{2\pi}\sum_{{\mathbf{k}}mn}\frac{J_{y}^{mn}({\mathbf{k}})J_{x}^{nm}({\mathbf{k}})}{\left(\varepsilon_{\mathbf{k}}^{m}-\varepsilon_{\mathbf{k}}^{n}\right)^{2}}\left(n_{\mathbf{k}}^{m}-n_{\mathbf{k}}^{n}\right)=\frac{e^{2}}{h},
\end{equation}
in the case of the integer Quantum Hall effect as in our system.

%The topological nature of the above expression becomes more transparent by noting that since for $l=x,y$:
%\begin{eqnarray*}
%0&=&{\partial \over \partial k_l} \langle m_{\mathbf k}|H({\mathbf k})| n_{\mathbf k}\rangle=\\
%& &\varepsilon^n_{\mathbf k} \langle {\partial \over \partial k_l}  m_{\mathbf k}|n_{\mathbf k}\rangle+\langle m_{\mathbf k}|J_l({\mathbf k})| n_{\mathbf k}\rangle+\varepsilon^m_{\mathbf k}\langle m_{\mathbf k}| {\partial \over \partial k_l} n_{\mathbf k}\rangle,\\
%0&=&{\partial \over \partial k_l} \langle m_{\mathbf k}| n_{\mathbf k}\rangle=
%\langle {\partial \over \partial k_l}  m_{\mathbf k}|n_{\mathbf k}\rangle+\langle\ m_{\mathbf k}| {\partial \over \partial k_l} n_{\mathbf k}\rangle,
%\end{eqnarray*}
%we have a general relationship
%\begin{equation}
%{J_l^{mn}({\mathbf k})\over \varepsilon^n_{\mathbf k}-\varepsilon^m_{\mathbf k} }=
%\langle m_{\mathbf k}| {\partial \over \partial k_l} n_{\mathbf k}\rangle
%\end{equation}
%and the fractional charge $q$ is related to an integral over the Berry curvature. More generally, however, an in particular if the magnetic structure has some net magnetization (for example by adding a Zeeman field) there are other nontopological contributions to the fractional charge, which leads to a total charge that can continuously change and take generally irrational values.

\section{Explicit form of Hamiltonian and current operators for the triangular and kagom\'e lattice}

\label{app:response}

All conductivities $\sigma_{xy}^{ab}$ can be computed from the general expression~\eqref{eq:cond}, but with the charge currents replaced by the appropriate charge or spin current $J_{xa}^{nm}({\mathbf{k}})$ and $J_{yb}^{nm}({\mathbf{k}})$.
This expression is applicable to translationally invariant systems where momentum $\bf k$ is a good quantum number. For a unit cell of $M$ sites, the momentum-space Hamiltonian ${\cal H}(\bf k)$ can be generically written as a $2M\times 2M$ matrix. (The factor of two accounts for electron spin.) For each momentum $\bf k$, we can diagonalize this $2M\times 2M$ Hamiltonian and obtain its eigenvalues and eigenvectors: ${\mathcal H}({\bf k})=\varepsilon^m_{\bf k}|m_{\bf k}\rangle \langle m_{\bf k}|$, $m=1\dots 2M$. The eigenvalues $\varepsilon^m_{\bf k}$ give the $2M$ energy bands. If we have symmetries, as in the triangular lattice case discussed below, these bands can be degenerate. The matrix elements $J_{xa}^{nm}({\mathbf{k}})$ can be constructed explicitly using the eigenvectors $|m_{\bf k}\rangle $ and $ |n_{\bf k}\rangle$, where $J({\bf k})$ is an appropriate current operator written as a $2M\times 2M$ matrix in the same basis as ${\mathcal H}({\bf k})$.
 
The only ingredients for computing $\sigma_{xy}^{ab}$ are then the $2M\times 2M$ Hamiltonian ${\mathcal H}(\bf k)$ and the corresponding $2M\times 2M$ charge and spin current operators $J^{x,y}_a({\bf k})$ for $a=0\dots 3$. With these ingredients, one can diagonalize ${\mathcal H}(\bf k)$ to obtain the eigenvalues and eigenvectors, use the eigenvectors to construct the matrix elements of the current operators, perform the sum over $m$ and $n$, and finally integrate the resulting expression over momenta ${\mathbf k}$ in the Brillouin zone.

Based on the above prescription, our main task is to write ${\mathcal H}({\bf k})$ and $J^{x,y}_a({\bf k})$. We first choose an explicit tetrahedral magnetic ordering represented as in Fig.~\ref{fig:trig}. As in the main text, we also assume an additional Zeeman field $h$ in the $z$ direction. We choose the following basis:
 \[\Psi^\dagger_{\mathbf{k}}=(c^\dagger_{1 \uparrow {\mathbf{k}}},c^\dagger_{1 \downarrow {\mathbf{k}}},c^\dagger_{2 \uparrow {\mathbf{k}}},c^\dagger_{2 \downarrow {\mathbf{k}}},c^\dagger_{3 \uparrow {\mathbf{k}}},c^\dagger_{3 \downarrow {\mathbf{k}}},c^\dagger_{4 \uparrow {\mathbf{k}}},c^\dagger_{4 \downarrow {\mathbf{k}}}),\]
to write the Hamiltonian as $H_T=\sum_{\mathbf{k}}\Psi^\dagger_{\mathbf{k}} {\cal H}_T(\mathbf{k})\Psi_{\mathbf{k}}$, where the subscript $T$ indicates the triangular lattice.
We can then write  
\begin{equation}
{\cal H}_T({\mathbf k})=  J
\left(\begin{array}{cccc}
\vec{S}_1\cdot{\boldsymbol\sigma}  &0 &0 &0 \\
0& \vec{S}_2\cdot{\boldsymbol\sigma} & 0 &0 \\
0 &0 &\vec{S}_3\cdot{\boldsymbol\sigma} &0 \\
0 &0 &0 &\vec{S}_4\cdot{\boldsymbol\sigma} 
\end{array} 
\right)+ {\cal E}_T({\mathbf k})\otimes \sigma^0 +h \openone \otimes \sigma^3,
\end{equation}
where 
\begin{equation}
{\cal E}_T ({\mathbf{k}})=\left(\begin{array}{cccc}
0 & \epsilon_{1}({\mathbf{k}}) & \epsilon_{3}({\mathbf{k}}) & \epsilon_{2}({\mathbf{k}})\\
\epsilon_{1}({\mathbf{k}}) & 0 & \epsilon_{2}({\mathbf{k}}) & \epsilon_{3}({\mathbf{k}})\\
\epsilon_{3}({\mathbf{k}}) & \epsilon_{2}({\mathbf{k}}) & 0 & \epsilon_{1}({\mathbf{k}})\\
\epsilon_{2}({\mathbf{k}}) & \epsilon_{3}({\mathbf{k}}) & \epsilon_{1}({\mathbf{k}}) & 0
\end{array}\right)
\end{equation}
with 
$\epsilon_{i}({\mathbf{k}}) \equiv -2t\cos({\mathbf{k}} \cdot{\mathbf{a}}_{i})$. 
Charge ($a=0$) and spin ($a=1,2,3$) current operators can then be simply written in the same basis as
\begin{equation}\label{eq:current}
J^x_a({\bf k})=\partial_{k_x}{\cal E}_T({\mathbf k})\otimes \sigma^a,\quad J^y_a({\bf k})=\partial_{k_y}{\cal E}_T({\mathbf k})\otimes \sigma^a.
\end{equation}

\begin{figure}[ht]
 \includegraphics[width=0.95\columnwidth]{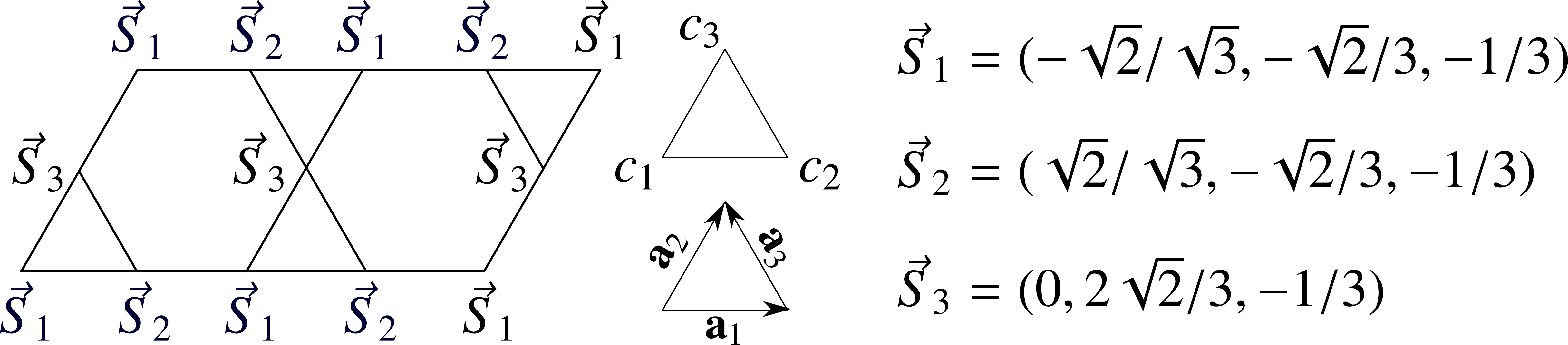}
\caption[]{An explicit chiral configuration of magnetic moments on the kagom\'e lattice. For site $i$ in sublattice $a=1\dots3$, ${\mathbf S}_i=\vec{S}_a$ as shown in the figure. The vectors ${\mathbf a}_i$ are the lattice vectors.}
\label{fig:kagome} 
\end{figure}

An almost identical expression can be written for the kagom\'e lattice by choosing explicit components for the magnetic moments and a basis $\Psi^\dagger_{\mathbf{k}}=(c^\dagger_{1 \uparrow {\mathbf{k}}},c^\dagger_{1 \downarrow {\mathbf{k}}},c^\dagger_{2 \uparrow {\mathbf{k}}},c^\dagger_{2 \downarrow {\mathbf{k}}},c^\dagger_{3 \uparrow {\mathbf{k}}},c^\dagger_{3 \downarrow {\mathbf{k}}})$ as shown in Fig.~\ref{fig:kagome}:
\begin{equation}
{\cal H}_K({\mathbf k})= {\cal E}_K({\mathbf k})\otimes \sigma^0 + J
\left(\begin{array}{cccc}
\vec{S}_1\cdot{\boldsymbol\sigma}  &0 &0  \\
0& \vec{S}_2\cdot{\boldsymbol\sigma} & 0  \\
0 &0 &\vec{S}_3\cdot{\boldsymbol\sigma} 
\end{array} 
\right)+ \openone \otimes \sigma^3,
\end{equation}
where 
\begin{equation}
{\cal E}_K ({\mathbf{k}})= \left(
\begin{array}{cccc}
0 & \epsilon_{1}({\mathbf{k}}) & \epsilon_{2}({\mathbf{k}}) \\
\epsilon_{1}({\mathbf{k}}) & 0 & \epsilon_{3}({\mathbf{k}}) 
\\
\epsilon_{2}({\mathbf{k}}) & \epsilon_{3}({\mathbf{k}}) & 0
\end{array}
\right).
\end{equation}

\section{Technical details of the Berry phase calculation}

\begin{figure}[ht]
\vspace{3mm}
 \includegraphics[width=0.9\columnwidth]{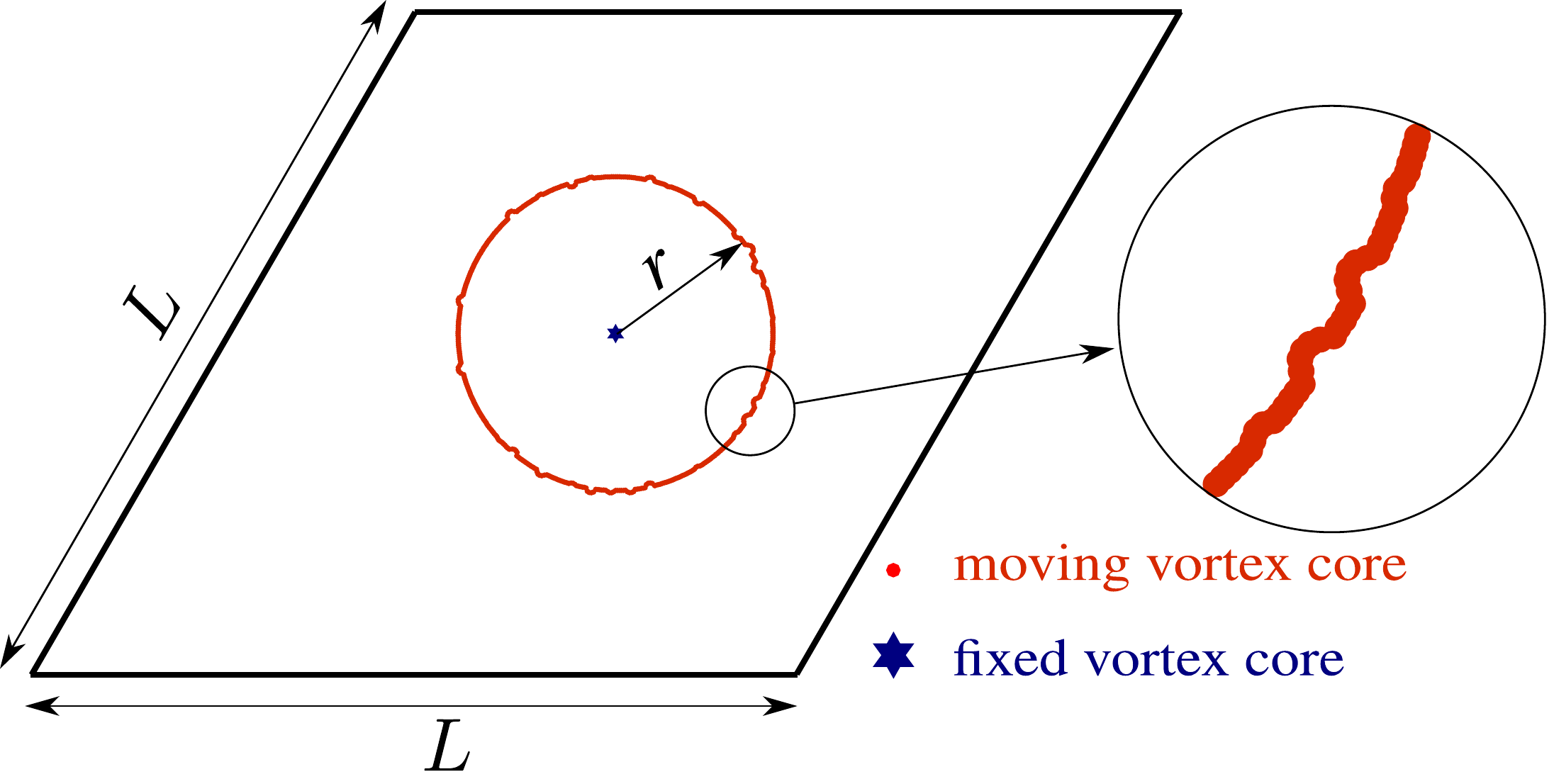} \caption[]{A typical path in a system of $L=40$ with radius $r=0.2L$. The
discretization points are distributed at uniform angular coordinate,
but the radial coordinate is shifted to maintain the minimal distance of $0.25$
(in units of lattice spacing) from the lattice sites. }
\label{fig:berry} 
\end{figure}
The idea is to simulate the adiabatic
motion of the vortex along a path by discretizing the path, and computing
the many-body ground-state wave function of the system for the vortex
core lying on each such discrete point along the path. Let us represent
these wave functions by $|\Psi_{i}\rangle$, with $i=1\dots N+1$,
and the periodic identification $|\Psi_{N+1}\rangle=|\Psi_{1}\rangle$.
We can then approximate the Berry phase $\oint\langle\Psi|\partial_{s}|\Psi\rangle ds$
by ~\cite{Ryu2009}
\[
\Phi=\arg\left[\prod_{i=1}^{N}\langle\Psi_{i}|\Psi_{i+1}\rangle\right],
\]
which is a convenient expression for numerical calculations as it
is explicitly gauge-invariant (each ground-state wave function obtained
from exact diagonalization has an arbitrary $U(1)$ phase; however, that phase obviously drops out from this expression). The only
underlying assumption for the validity of the above expression is
$1-|\langle\Psi_{i}|\Psi_{i+1}\rangle|\ll1$, which requires having
close discretization points so the overlaps $|\langle\Psi_{i}|\Psi_{i+1}\rangle|$
are close to one.

For the case of hard-core vortices in a noncoplanar magnet, the wave
functions can exhibit violent changes if the ``moving" vortex core
approaches very near a magnetic moment (i.e., a lattice site). Thus, in
order for the Berry phase to converge faster, it is important to use more discretization points in regions
of the path close to a lattice site, or, alternatively distort the
path in the vicinity of sites so as to maintain a minimum distance from them.
Here, we choose the latter approach with a typical path shown in Fig.~\ref{fig:berry}.
We first considered equally spaced discretization points (core of the
moving vortex) on a circle centered at the other (fixed) vortex. For
each discretization point we then find the distance to the closest
lattice site and, if necessary, distort the path by shifting the point
along the radial direction to the distance of $1/4$ away from the site. This simple trick results in large $|\langle\Psi_{i}|\Psi_{i+1}\rangle|$
overlaps.

To obtain reliable Berry phases from such numerics, we must have convergence in the number of discretization points (which requires large overlaps
of consecutive wave functions). Also, due the presence of edge modes
for open boundary conditions, we must be careful about level crossings
at the Fermi level. We work at a chemical potential inside the gap
but as the energy of the intragap states depends on microscopic details
(like the current location of the ``moving" vortex core inside a unit cell), for some trajectories
at fixed particle number, the last filled level can switch from an
edge mode to a bound state, rendering the results unreliable. This
can be easily checked a posteriori by keeping track of the energies
of the intragap levels as the vortex
moves along the path.

\bibliography{itinerant_magnets}{}

\end{document}